\documentclass[aps,pra,superscriptaddress,longbibliography]{revtex4-2}

\usepackage[utf8]{inputenc}
\usepackage{amsmath}
\usepackage{graphicx}
\usepackage{float}
\usepackage{afterpage}
\usepackage{siunitx}
\usepackage{gensymb}
\usepackage[version=4]{mhchem}
\usepackage[colorlinks,citecolor=red,urlcolor=blue,bookmarks=false,hypertexnames=true]{hyperref}
\usepackage[capitalize]{cleveref}

\hypersetup{
  linkcolor  = blue,
  citecolor  = blue,
  urlcolor   = blue,
  colorlinks = true,
}

\newcommand{\suppfigref}[1]{Supplementary Fig.~\ref*{#1}}

\newcommand{\suppsecref}[1]{Supplementary Section~\ref*{#1}}
\input{SIrefs}

\makeatletter
\renewcommand*{\@fnsymbol}[1]{\ensuremath{\ast}}
\makeatother

\begin{document}

\title{Thermal history controls the optoelectronic response of lead halide perovskites through structure and dynamics}

\author{Milos Dubajic}
\email{milos.dubajic@hotmail.com}
\affiliation{Department of Chemical Engineering and Biotechnology, University of Cambridge, Philippa Fawcett Drive, Cambridge, CB3 0AS, UK}

\author{Xia Liang}
\affiliation{Department of Materials, Imperial College London, London SW7 2AZ, United Kingdom}
\affiliation{School of Energy and Chemical Engineering, Ulsan National Institute of Science and Technology (UNIST), Ulsan 44919, Korea}

\author{Johan Klarbring}
\affiliation{%
Department of Physics, Chemistry and Biology (IFM),
Link\"{o}ping University, SE-581 83, Link\"{o}ping, Sweden
}%

\author{Yang Lu}
\affiliation{Department of Chemical Engineering and Biotechnology, University of Cambridge, Philippa Fawcett Drive, Cambridge, CB3 0AS, UK}
\affiliation{Department of Physics, Cavendish Laboratory, University of Cambridge, JJ Thomson Avenue, Cambridge, CB3 0HE, UK}

\author{Thomas A. Selby}
\affiliation{Department of Chemical Engineering and Biotechnology, University of Cambridge, Philippa Fawcett Drive, Cambridge, CB3 0AS, UK}

\author{Erik Fransson}
\affiliation{ Department of Physics, Chalmers University
of Technology, SE-41296 Gothenburg, Sweden}

\author{Philippe Holzhey}
\affiliation{Helmholtz-Zentrum Berlin für Materialien und Energie, Hahn-Meitner-Platz 1, 12489
Berlin, Germany}
\affiliation{Clarendon Laboratory, Department of Physics, University of Oxford, Parks Road, Oxford OX1 3PU, UK.}

\author{Benjamin M Gallant}
\affiliation{Helmholtz-Zentrum Berlin für Materialien und Energie, Hahn-Meitner-Platz 1, 12489 Berlin, Germany}
\affiliation{Clarendon Laboratory, Department of Physics, University of Oxford, Parks Road, Oxford OX1 3PU, UK.}

\author{Qichun Gu}
\affiliation{Department of Chemical Engineering and Biotechnology, University of Cambridge, Philippa Fawcett Drive, Cambridge, CB3 0AS, UK}

\author{Ganbaatar Tumen-Ulzii}
\affiliation{Department of Chemical Engineering and Biotechnology, University of Cambridge, Philippa Fawcett Drive, Cambridge, CB3 0AS, UK}

\author{Khasim Saheb Bayikadi}
\affiliation{Department of Chemistry, Northwestern University, Evanston, IL, 60208 USA}

\author{Isaiah Gilley}
\affiliation{Department of Chemistry, Northwestern University, Evanston, IL, 60208 USA}

\author{{Martin v. Zimmermann}}
\affiliation{    Deutsches Elektronen-Synchrotron DESY, Notkestraße 85, 22607, Hamburg, Germany}

\author{ Christian Orr}
\affiliation{Diamond Light Source, UK}

\author{Chwenhaw Liao}
\affiliation{School of Physics, The University of Sydney, Sydney, New South Wales, Australia}

\author{Josh S. Moon}
\affiliation{Department of Materials Science and
Engineering, Monash University, Clayton, VIC 3800,
Australia}

\author{Jacek Jasieniak}
\affiliation{Department of Materials Science and
Engineering, Monash University, Clayton, VIC 3800,
Australia}

\author{Makhsud Saidaminov}
\affiliation{Department of Chemistry, University of Victoria, Victoria, British Columbia V8P5C2, Canada}

\author{Michael P. Nielsen}
\affiliation{School of Photovoltaic \& Renewable Engineering, UNSW Sydney, Kensington 2052, Australia}

\author{Tom Wu}
\affiliation{Department of Applied Physics, The Hong Kong Polytechnic University, Hong Kong, China}

\author{Stephen P. Bremner}
\affiliation{School of Photovoltaic \& Renewable Engineering, UNSW Sydney, Kensington 2052, Australia}

\author{Anita Ho-Baillie}
\affiliation{School of Physics, The University of Sydney, Sydney, New South Wales, Australia}

\author{Julia Wiktor}
\affiliation{ Department of Physics, Chalmers University
of Technology, SE-41296 Gothenburg, Sweden}

\author{Paul Erhart}
\affiliation{ Department of Physics, Chalmers University
of Technology, SE-41296 Gothenburg, Sweden}

\author{Steve Albrecht}
\affiliation{Helmholtz-Zentrum Berlin für Materialien und Energie, Hahn-Meitner-Platz 1, 12489 Berlin, Germany}

\author{Mercouri Kanatzidis}
\affiliation{Department of Chemistry, Northwestern University, Evanston, IL, 60208 USA}

\author{Henry J Snaith}
\affiliation{Clarendon Laboratory, Department of Physics, University of Oxford, Parks Road, Oxford OX1 3PU, UK.}

\author{Aron Walsh}
\affiliation{Department of Materials, Imperial College London, London SW7 2AZ, United Kingdom}

\author{Samuel D. Stranks}
\email{sds65@cam.ac.uk}
\affiliation{Department of Chemical Engineering and Biotechnology, University of Cambridge, Philippa Fawcett Drive, Cambridge, CB3 0AS, UK}
\affiliation{Department of Physics, Cavendish Laboratory, University of Cambridge, JJ Thomson Avenue, Cambridge, CB3 0HE, UK}

\begin{abstract}

Lead halide perovskites have emerged as a promising class of optoelectronic materials for photovoltaics, light emission and detection. Their efficiencies in PV now approach the detailed-balance limit, leaving stability as the principal barrier. The intrinsic instabilities studied to date centre on ionic motion, framed within a fixed, homogeneous crystal lattice. Here we identify a further source of intrinsic structural instability, hidden in the unique lattice dynamics properties. Mapping caesium, methylammonium and formamidinium-based compositions with Cl, Br, I and mixed-halide X-sites through all accessible phases, using single crystal X-ray and neutron diffuse scattering, machine-learning-assisted molecular dynamics simulations, a phenomenological octahedral tilt model and temperature-dependent hyperspectral photoluminescence, we discover that nearly every composition hosts equilibrium local structural fluctuations: dynamic nanodomains of correlated octahedral tilts, a few nanometres in size, that locally break the crystallographic symmetry. Three complementary levers control them. The A-site cation sets their symmetry, shape and anisotropy, from sparse, isotropic and tetragonal in formamidinium-based compositions to dense, anisotropic and orthorhombic in nominally cubic caesium-based compositions, the most locally disordered of all compositions studied. The halide controls the dynamic disorder and the phase-transition sequence. Thermal history plays a key role: different ramp rates drive nominally identical compositions into distinct crystallographic phases, each with its own hidden local order. We demonstrate the consequences in \ce{MAPbI3}, where the heating rate alone changes the photoluminescence quantum efficiency across its phase transition. Because these transitions lie within the operating temperature range of perovskite devices, from terrestrial thermal cycling to the extremes of space, thermal history may shape the local structure, and hence the optoelectronic response, during fabrication and throughout operation. This acute sensitivity helps explain why perovskite performance depends so strongly on processing and establishes thermal history, alongside composition, as a design variable.

\end{abstract}

\maketitle
\newpage

Lead halide perovskites (LHPs) have rapidly established themselves as low-cost, high-performance semiconductors for solar cells, light-emitting diodes and radiation detection \cite{sutherland_perovskite_2016, sakhatskyi_stable_2023, jena_halide_2019}.
With power conversion efficiencies in PV applications approaching the detailed-balance limit \cite{PVtables_v6}, the principal barrier to their deployment is stability \cite{boyd_degradation_2019}. The instabilities identified so far centre on ionic motion, which is regarded as intrinsic in that it cannot be eliminated by encapsulation \cite{yuan_ion_2016}: halide and cation migration through the soft lattice produces current-voltage hysteresis, drives light-induced halide segregation in mixed-halide absorbers \cite{hoke_segregation_2015} and supplies the mobile species for electrochemical degradation at interfaces and electrodes \cite{boyd_degradation_2019}. Thus, the crystal that hosts the mobile ions is treated as a fixed, homogeneous lattice, the time-averaged structure resolved by conventional diffraction. However, in these materials, this simplified picture does not apply and many functional properties of LHPs cannot be understood from the average structure alone.

~

This is not unique to LHPs: across many functional solids, a hidden local order, distinct from the average structure and built from short-range atomic correlations, governs key properties, reshaping the electronic and phononic band structure \cite{roth_tuning_2023} and accounting for otherwise unexplained behaviour in battery electrodes \cite{krogstad_reciprocal_2020,simonov_hidden_2020}, relaxor ferroelectrics \cite{krogstad_relation_2018} and cuprate superconductors \cite{pelc_unconventional_2022}. It has been established that in certain LHPs this hidden order takes the form of dynamic nanodomains of correlated octahedral tilts, a few nanometres across, that fluctuate within the nominally high-symmetry phases and are directly inferable from characteristic features in macroscopic diffuse scattering results \cite{weadock_nature_2023,dubajic_dynamic_2025}.
We can use single crystal X-ray diffraction to separate sharp Bragg peaks, which report the long-range-ordered average structure, from diffuse scattering: weaker, broader intensity (usually at high symmetry reciprocal-space positions, e.g. R, M and X) which in LHPs is set by the symmetry of octahedral tilting, whose finite width corresponds to short-range rather than long-range tilt correlations (\suppsecref{sup_ssec:Diffuse_Model}). In certain LHPs this diffuse signal is quasi-elastic (QEDS) and arises from nanodomains, nanometre-scale regions of correlated octahedral tilts that fluctuate within the nominally high-symmetry phases, an interpretation supported both by phenomenological modelling  and molecular dynamics (MD) simulations \cite{weadock_nature_2023,dubajic_dynamic_2025}. These nanodomains are shown to be dynamic, forming and dissolving on picosecond timescales \cite{dubajic_dynamic_2025,weadock_nature_2023}.

~

The growing evidence suggests that these nanodomains govern macroscopic properties across a wide range of applications. The highly anharmonic lattice potential energy surface rooted in the local disorder drive phonon-mediated electronic wavefunction localization in perovskite quantum dots, enhancing single-photon purity at room temperature \cite{feld_phonon_2026}. Strain gradients at ferroelastic nanodomain walls generate flexoelectric polarization that enables charge separation and long-lived photocurrents in nominally cubic crystals \cite{rak_flexoelectric_2026}. The same anharmonicity underpins the anomalous thermal expansion that may limit the thermo-mechanical stability of perovskite solar cells \cite{steele_anharmonic_2026}. These recent findings suggest that understanding, and ultimately controlling, the nanodomain landscape is essential for future progress of LHP materials.

~

We recently showed that in the cubic phases of \ce{MAPbBr3} and \ce{FAPbBr3}, the A-site cation dictates the symmetry, shape and dynamic disorder of such nanodomains: methylammonium (MA)-based perovskites form densely packed, anisotropic nanodomains, whereas formamidinium (FA)-based perovskites develop sparse, isotropic nanodomains, with measurable consequences for the electronic disorder, (as quantified by Urbach energy) and photoluminescence (PL) quantum efficiency \cite{dubajic_dynamic_2025}.
This established the link in two bromide compositions but left broader questions open: whether such nanodomains are universal across the LHP family, what determines their symmetry, anisotropy and the resulting dynamic disorder, and whether they can be deliberately engineered.
Answering these questions requires exploring the changes in local structure across the full compositional space, disentangling the roles of the A- and X-site sublattices from those of the processing route, and revealing how sensitively nanodomain formation responds to thermal processing conditions.

~

Here, we achieve this by combining single crystal X-ray and neutron diffuse scattering with machine-learning-assisted MD simulations and a phenomenological octahedral tilt model, surveying \ce{Cs+}-, \ce{MA+}-, and \ce{FA+}-based perovskites with \ce{Cl-}, \ce{Br-}, \ce{I-}, and mixed-halide $X$-sites across a wide temperature range encompassing all crystallographic phases.
We show that the symmetry, shape and dynamic disorder of the nanodomains can be controlled through three complementary levers, the A-site cation, the halide and the thermal history, with thermal history alone driving a single composition into different local structures depending on the cooling rate. Across all compositions studied, we account for every feature of the diffuse scattering and identify its origin. In the Cs-based perovskites we additionally identify a signature of correlated Cs and Pb displacements, an off-centering debated for years \cite{laurita_chemical_2017}, which we show to be consistent with arising from phonon branches with energies between 2 and 3 meV. By controlling how the nanodomains evolve through a phase transition, we gain direct control of the optoelectronic response: in \ce{MAPbI3}, the heating rate tunes the PL intensity, and hence the PL quantum efficiency (PLQY), across the phase transition. It has been known that the structural identity of an LHP is not fixed by composition alone, but is co-determined by its processing history. We show here that the thermal history is an additional degree of freedom that sets both the average phase and the local order, providing design rules for engineering the optoelectronic response of these materials.

\subsection*{Compositional engineering reveals diverse local structural landscape}

Three distinct classes of local structure emerge across the LHP series studied, governed primarily by the A-site cation.
\cref{fig:Fig1} presents the experimental scattering function $S(\mathbf{q})$ across $\text{HK}1.5$ reciprocal space planes for Cs-, MA- and FA-based LHPs with Cl, Br, I and mixed-halide X-sites, organised by average phase: cubic (\cref{fig:Fig1}a, b, c), tetragonal (\cref{fig:Fig1}d, e), and incommensurate (\cref{fig:Fig1}f), with an overview of the composition-dependent local structural landscape in \cref{fig:Fig1}h.
Real-space schematics illustrating each local structure configuration using Glazer grid notation are provided in \suppsecref{sup_sec:Schematics}.

~

In the average cubic ($Pm\bar{3}m$) phases, we identify three distinct types of local structure governed primarily by the A-site cation (\cref{fig:Fig1}a, b, c).
Cs-based perovskites (\ce{CsPbBr3} and \ce{CsPbCl3}) exhibit diffuse scattering peaks at both the $R$ and $M$ high-symmetry points, consistent with local, dynamic orthorhombic ($a^{-}a^{-}c^{+}$) nanodomains (\cref{fig:Fig1}a,  \suppfigref{S-fig:Schematic_Case3}).
These nanodomains form pancake-shaped correlations along all three octahedral tilt directions, twinned with six components arising from the three equivalent orientations of the orthorhombic distortion within the cubic matrix (\cref{fig:Fig2} d).
In addition to the rod-like diffuse scattering from octahedral tilts, Cs-based compositions display a distinctive flower-like diffuse scattering pattern superimposed on the rods, which we attribute to correlated A-site (Cs) and B-site (Pb) motions (\cref{fig:Fig2} e) \cite{kopecky_diffuse_2012}.

~

MA-based perovskites (\ce{MAPbBr3}, \ce{MAPbI3} and mixed \ce{MAPbI_{x}Br_{3-x}}) display diffuse peaks exclusively at the $R$ points, indicating local anti-phase tetragonal ($a^{0}a^{0}c^{-}$, $I4/mcm$) nanodomains with pancake-shaped morphology (\cref{fig:Fig1}b; \suppfigref{S-fig:Schematic_Case2}).
FA-based perovskites (\ce{FAPbBr3} and \ce{FAPbI3}) show diffuse scattering at the $M$ points, characteristic of local, spherical, in-phase tetragonal ($a^{0}a^{0}c^{+}$, $P4/mbm$) nanodomains (\cref{fig:Fig1}c and \suppfigref{S-fig:Schematic_Case1}).
The triple cation (TC) perovskite \ce{Cs_{0.05}(MA_{0.17}FA_{0.83})_{0.95}Pb(I_{0.83}Br_{0.17})3} \cite{saliba_cesium-containing_2016} exhibits the same $M$-point diffuse scattering as the FA-based end members, indicating that the FA-dominated A-site governs the local structure symmetry even in the alloyed composition.

~

Lowering the temperature in MA systems induces ferroelastic phase transitions into tetragonal phases. The symmetry of the dynamic nanodomains matches that of the average tetragonal phase (locally $a^{0}a^{0}c^{-}$ within average $a^{0}a^{0}c^{-}$; \cref{fig:Fig1}e and \suppfigref{S-fig:Schematic_Case5}).
In Cs-based systems, the local dynamic structure retains its orthorhombic ($a^{-}a^{-}c^{+}$) character even within the average tetragonal ($a^{0}a^{0}c^{+}$) phase (\cref{fig:Fig1}d; \suppfigref{S-fig:Schematic_Case4}). 
At low temperatures, certain compositions (\ce{MAPbBr3}, \ce{FAPbBr3} and \ce{MAPbIBr2}) develop satellite reflections that cannot be indexed as rational multiples of the reciprocal-lattice vectors (\cref{fig:Fig1}f), while the remaining compositions retain the $Pnma$ or $Im\bar{3}$ phases. 
These satellite reflections are unambiguous evidence of an incommensurate modulation, which arises when a periodic distortion superimposed on the lattice has a period that is not a rational multiple of the lattice spacing. In halide perovskites this distortion is a sinusoidal modulation of the octahedral tilts \cite{guo_interplay_2017,wiedemann_hybrid_2021,li_dual_2021}.

~

Cs-based perovskites form pancake-shaped nanodomains along all three tilt directions simultaneously ($\xi_{\parallel} \approx 35$~\AA\ vs.~$\xi_{\perp} \approx 7$~\AA, experimentally determined in \ce{CsPbBr3} in the cubic phase at 425 K).
These systems exhibit correlated octahedral tilts in both the cubic and tetragonal phases (\suppfigref{S-fig:Schematic_Case3}, \suppfigref{S-fig:Schematic_Case4}), and the distinctive flower-like diffuse scattering pattern, observed in every phase of the Cs perovskites studied, indicates that correlated Cs and Pb motions persist throughout.
In the orthorhombic $Pnma$ phase, however, the diffuse scattering signatures are absent, thus we infer that the octahedral tilts become statically long-range ordered and without the presence of any dynamic nanodomains (\cref{fig:Fig2}f).

~

MA-based perovskites exhibit local structure in both the cubic and tetragonal phases.
The symmetry of the local structure in the tetragonal phase matches the average phase (\suppfigref{S-fig:Schematic_Case2}, \suppfigref{S-fig:Schematic_Case5}).
Certain MA compositions, such as \ce{MAPbI3} and \ce{MAPbCl3}, do not exhibit local structure (dynamic nanodomains) in the orthorhombic phase, where only superstructure Bragg reflections remain (\suppfigref{S-fig:HK15_Pure_Halides}), while others, such as \ce{MAPbBr3}, develop incommensurate modulations at lower temperatures (\suppfigref{S-fig:MAPbBr3_P211}).
FA-based perovskites undergo a cubic-to-cubic phase transition ($Pm\bar{3}m \rightarrow Im\bar{3}$, $a^{+}a^{+}a^{+}$) that removes the nanodomains entirely; in the resulting cubic phase, no quasi-elastic diffuse scattering (QEDS) is observed.
The correlation lengths extracted from the diffuse scattering peaks are comparable across the different compositions within each A-cation family (\suppfigref{S-fig:Correlation_Lengths}), with clear nanodomain anisotropy trends.
\begin{figure}[H]
    \centering
    \includegraphics[width=1.05\textwidth]{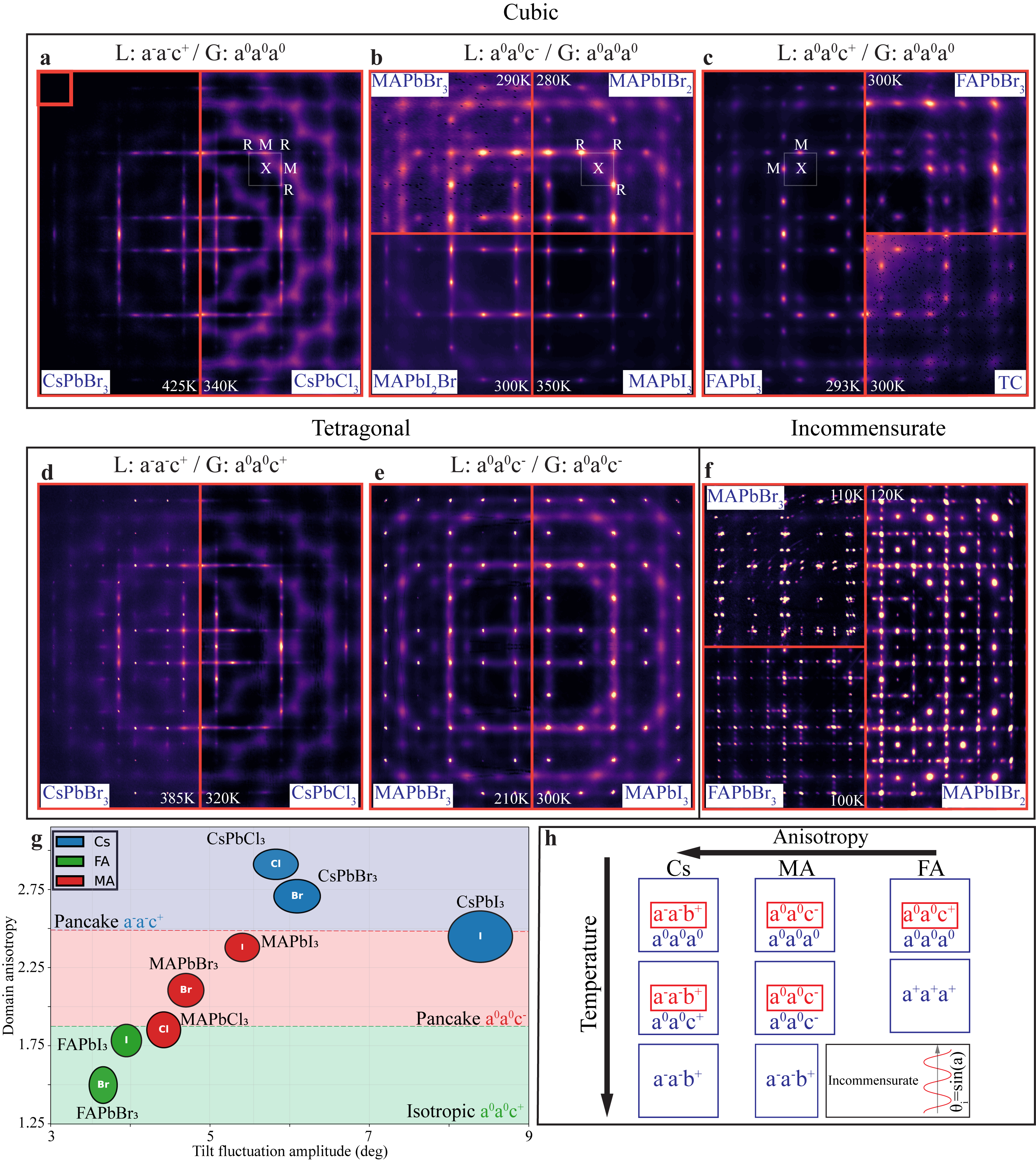}
\end{figure}

\afterpage{%
\begin{figure}[t]
\caption{{\bf Diverse local structural landscapes in LHPs revealed by X-ray diffuse scattering and MD simulations.}
Experimental X-ray scattering function $S(\mathbf{q})$ across $\text{HK}1.5$ reciprocal space planes for various LHP compositions and temperatures. Each sub-panel shows measurements at different temperatures (indicated in the corners), with compositions labelled.
\textbf{a, b, c}~Local structure in average cubic ($Pm\bar{3}m$) phases. \textbf{a}: local orthorhombic ($a^{-}a^{-}c^{+}$) nanodomains in \ce{CsPbBr3} and \ce{CsPbCl3}, with diffuse scattering at both $R$ and $M$ points. The red square in the upper-left corner serves as a scale bar, representing $1 \times 1\text{ r.l.u.}$ along the respective reciprocal-space dimensions. \textbf{b}: local anti-phase tetragonal ($a^{0}a^{0}c^{-}$) nanodomains in \ce{MAPbBr3}, \ce{MAPbI_{x}Br_{3-x}} and \ce{MAPbI3}, with diffuse peaks at $R$ points. \textbf{c}: local in-phase tetragonal ($a^{0}a^{0}c^{+}$) nanodomains in \ce{FAPbBr3} and \ce{FAPbI3}, with diffuse peaks at $M$ points. TC denotes the triple cation perovskite \ce{Cs_{0.05}(MA_{0.17}FA_{0.83})_{0.95}Pb(I_{0.83}Br_{0.17})3}.
\textbf{d, e}~Local structure in average tetragonal phases. \textbf{d}: local orthorhombic ($a^{-}a^{-}c^{+}$) in average tetragonal ($a^{0}a^{0}c^{+}$) for \ce{CsPbBr3} and \ce{CsPbCl3}. \textbf{e}: local anti-phase tetragonal ($a^{0}a^{0}c^{-}$) matching the average tetragonal symmetry ($a^{0}a^{0}c^{-}$) for \ce{MAPbBr3} and \ce{MAPbI3}.
\textbf{f}~Incommensurate structural modulations observed in \ce{MAPbBr3}, \ce{FAPbBr3} and \ce{MAPbIBr2}, characterised by satellite reflections from sinusoidal modulation of octahedral tilts.
\textbf{g}~Disorder-anisotropy diagram of the cubic LHPs computed from MD trajectories. The $x$-axis is the per-axis rms octahedral tilt-fluctuation amplitude $\sigma_\theta$, a measure of dynamic disorder, and the $y$-axis is the domain anisotropy $\eta_\xi = \xi_{\parallel}/\xi_{\perp}$ of the dynamic tilt-correlation-length matrix. Each composition is drawn as an ellipse whose aspect ratio equals $\eta_\xi$ and whose area scales with $\sigma_\theta$, coloured by A-cation family (blue: Cs, red: MA, green: FA). The two measures are positively correlated  and separate the compositions into three regimes: near-isotropic in-phase nanodomains ($a^{0}a^{0}c^{+}$, FA), anisotropic anti-phase  ($a^{0}a^{0}c^{-}$, MA), and the most anisotropic, $Pnma$-like nanodomains ($a^{-}a^{-}c^{+}$, Cs).
\textbf{h}~Overview of the composition-dependent local structural landscape derived from MD simulations and experiments for Cs-, MA- and FA-based systems as a function of temperature. Blue boxes denote the average structure and red boxes denote the local structure, each labelled with the corresponding Glazer notation. Incommensurate phases with sinusoidal modulation of octahedral tilt angle are present in MA and FA systems at the lowest temperatures.  The anisotropy of the dynamic nanodomains increases from FA  through MA  to Cs, as indicated by the arrow.}
    \label{fig:Fig1}
\end{figure}%
}
\newpage

Here, these dynamic nanodomains represent the spatial envelopes over which transient octahedral tilt fluctuations are correlated. FA-based perovskites form near-isotropic nanodomains, while MA- and Cs-based perovskites form anisotropic, pancake-shaped nanodomains, where the correlation lengths are highly direction-dependent. The strength of the QEDS serves as a rough estimate of the dynamic disorder, and we find that this disorder increases from Cl to Br to I.
We independently verified these cation-family trends with MD simulations (\cref{fig:Fig1}g). For each cubic composition, we quantify the dynamic disorder as the width $\sigma_\theta$ of the Gaussian distribution of octahedral tilt angles, averaged over the three pseudocubic axes, so that a larger width corresponds to a more disordered lattice containing a higher fraction of large-amplitude tilt nanodomains. This measure and the anisotropy of the nanodomains together separate the compositions into three regimes (\cref{fig:Fig1}g): near-isotropic in-phase nanodomains in the FA systems, anisotropic anti-phase pancakes in the MA systems, and the most anisotropic, $Pnma$-like nanodomains in the Cs systems. This MD-derived ordering reproduces the cation-family trend inferred from the experimental diffuse scattering and summarised in \cref{fig:Fig1}h (\suppsecref{sup_ssec:CrossMaterial_anisotropy}).

~

Taken together, the compositional survey establishes composition as the primary handle on the nanodomain landscape: the A-site cation sets the symmetry and anisotropy of the nanodomains, from near-isotropic (FA), through anisotropic pancakes (MA), to the most anisotropic, lowest-symmetry orthorhombic (Cs), while the halide sets the degree of dynamic disorder, which increases from Cl through Br to I.
The landscape therefore runs between two extremes: FA- and bromide-rich compositions such as \ce{FAPbBr3}, with the least disordered, isotropic nanodomains, and Cs- and iodide-rich compositions, with the most disordered and anisotropic, lowest-symmetry nanodomains.

\subsection*{Local orthorhombic nanodomains and correlated cation displacements in \ce{CsPbBr3}}

\begin{figure}[H]
    \centering
    \includegraphics[width=0.97\textwidth]{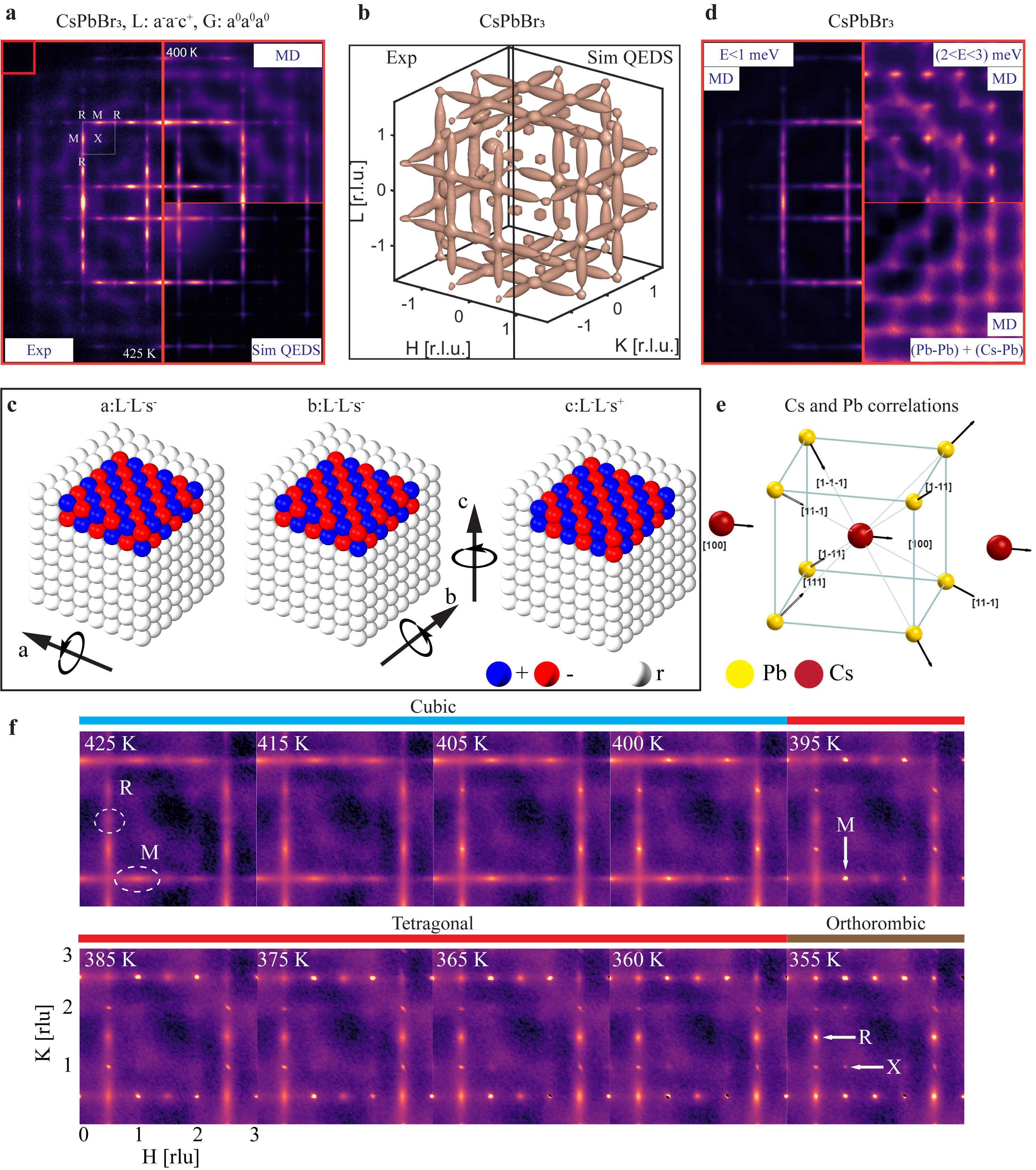}
\caption{{\bf Local orthorhombic nanodomains arise from overdamped soft phonon modes while  correlated cation displacements emerge from thermal diffuse scattering in cubic \ce{CsPbBr3}.}
\textbf{a,}~Left: experimental $S(\mathbf{q})$ in the $\text{HK}1.5$ plane at \SI{425}{\kelvin} (average cubic phase). The red rectangle indicates a $1 \times 1$~r.l.u.\ region. Right, top: MD simulation at \SI{400}{\kelvin}. Right, bottom: phenomenological model simulation of QEDS.
\textbf{b,}~Three-dimensional $S(\mathbf{q})$ volume from experiment (left) and the phenomenological QEDS model (right), demonstrating close agreement.
\textbf{c,}~Real-space visualisation of the $Pnma$ nanodomains, viewed along the three pseudocubic directions. Each octahedron is represented by a sphere coloured by the tilting angle along the relevant axis (blue, positive; red, negative; grey, near-zero). The modified Glazer notation indicates the correlation type along each direction: $\text{s}$ = short-range, $\text{L}$ = long-range.
\textbf{d,}~Energy-resolved $S(\mathbf{q},\omega)$ from MD simulations of \ce{CsPbBr3}. Left: QEDS window ($E<1$~meV), in which only the $R$- and $M$-point rods appear. Top right: the 2--3~meV window, in which the flower-like pattern appears and the rods are absent. Bottom right: the sum of the Pb--Pb and Cs--Pb partial structure factors, which reproduces the flower pattern.
\textbf{e,}~Schematic of correlated Cs and Pb displacements: Cs atoms (red) are displaced along $\langle 100 \rangle$ directions and Pb atoms (yellow) along $\langle 111 \rangle$ directions, producing the flower-like diffuse scattering pattern.
\textbf{f,}~Temperature evolution of $S(\mathbf{q})$ at $\text{HK}1.5$ across the cubic, tetragonal and orthorhombic phases of \ce{CsPbBr3}. $M$-point Bragg peaks emerge at the cubic$\rightarrow$tetragonal transition (by \SI{395}{\kelvin}) and $R$- and $X$-point Bragg peaks at the tetragonal$\rightarrow$orthorhombic transition (by \SI{355}{\kelvin}). The flower-like diffuse scattering from Cs--Pb correlations persists across all phases of \ce{CsPbBr3}.}
    \label{fig:Fig2}
\end{figure}

To determine the properties of the local structure in all-inorganic LHPs, a phenomenological octahedral tilt model is applied to experimental diffuse scattering data of cubic \ce{CsPbBr3} and \ce{CsPbCl3}.
Diffuse scattering is observed at both the $R$ and $M$ points, with ellipsoid (rod-like) peaks at $M$ and near-spherical features at $R$ (\cref{fig:Fig2}a).
We can reproduce the data using an orthorhombic $Pnma$ local structure with anisotropic, pancake-shaped nanodomains described by only two correlation lengths: $\xi_{\perp} = 6.5$~\AA\ (normal to the pancake plane) and $\xi_{\parallel} = 35.1$~\AA \,  (\suppfigref{S-fig:CsPbBr3_PhenomFit_2D}).
We tested alternative models that imply different shape and symmetry of nandomains all of which cannot fit the experimental data (\suppfigref{S-fig:Alternative_Fits}).
The experimental data and the QEDS fit are shown in \cref{fig:Fig2}a, and the corresponding three-dimensional reciprocal-space volumes for experiment and the phenomenological model are compared in \cref{fig:Fig2}b, demonstrating close agreement.
MD simulations also reproduce the experimental trends (\cref{fig:Fig2}a, top right).
By contrast, real-space analysis of the MD trajectories is complicated by local correlations along all three tilt directions ($a^{-}a^{-}c^{+}$), combined with the presence of six twin variants.
The phenomenological model therefore provides a practical, computationally efficient route to identifying hidden local order in perovskite structures.

~

We further note that the twinning laws used for the cubic-phase nanodomains are the same as those in the ferroelastic $Pnma$ phase of \ce{CsPbBr3} and \ce{CsPbCl3}.
Consistent with this, we confirm that the room-temperature phase is $Pnma$ orthorhombic (as opposed to different interpretations recently suggested in the literature \cite{liu_high-resolution_2021,cspbbr3_diff_phase}), as non-merohedral twinning in these ferroelastic materials lowers the apparent symmetry when sampling over many twins, producing  more superstructure peaks than expected for a single $Pnma$ domain (\suppfigref{S-fig:CsPbCl3_Twinning_Combined}).
\cref{fig:Fig2}c presents the schematic of the $Pnma$-like nanodomains in real space.
Unlike the tetragonal nanodomains in \ce{MAPbBr3} and \ce{FAPbBr3} \cite{dubajic_dynamic_2025}, which are correlated along a single tilt direction, the orthorhombic nanodomains in \ce{CsPbBr3} exhibit tilt correlations along all three directions simultaneously (in both space and time): short-range along $c$ and longer-range along $a$ and $b$, consistent with $a^{-}a^{-}c^{+}$ local nanodomains. 
This finding aligns with Baldwin et al., who proposed that the local structure in cubic \ce{CsPbI3} is of lower symmetry than tetragonal \cite{baldwin_dynamic_2024}.
We previously showed that FA molecules act as out-of-plane tilt-coupling elements that drive the isotropy of the dynamic nanodomain correlation lengths in LHPs \cite{dubajic_dynamic_2025}. 
In Cs-based compositions, we instead observe pancake-like octahedral tilt nanodomains along all three tilt directions, resembling the  local orthorhombic phase. We previously showed that the anharmonic potential energy surface of the corner-sharing octahedral network inherently favours strong in-plane over out-of-plane tilt correlations, so planar, pancake-like nanodomains are the intrinsic outcome of the anharmonicity; hydrogen bonding between the A-site cation and the halide sublattice can counteract this inherent anisotropy by coupling adjacent octahedral layers out-of-plane, extending correlations in that direction and making the nanodomains more isotropic \cite{dubajic_dynamic_2025}. FA forms the strongest, most directional hydrogen bonds and yields the largest increase in isotropy, MA's weaker, less directional bonding has a smaller effect, and Cs, with no hydrogen-bonding capacity at all, leaves the intrinsic anisotropy unmodified, giving the thinnest nanodomains and the largest anisotropy of the compositions studied.

~

Beyond the octahedral tilt nanodomains, a distinctive flower-like diffuse scattering pattern is observed in \ce{CsPbBr3}, \ce{CsPbCl3} and \ce{CsPbI3} in all phases (\cref{fig:Fig1}a; \suppfigref{S-fig:Flower_AllPhases}), which we attribute to correlated Cs and Pb displacements (\cref{fig:Fig2}e).
An analogous flower-like diffuse scattering has been reported in \ce{SrTiO3}, where it arises from correlated displacements of the A- and B-site cations \cite{kopecky_diffuse_2012}.
In this picture, Cs atoms are correlated along the $\langle 100 \rangle$ directions while Pb atoms are correlated along the $\langle 111 \rangle$ directions.
These Cs--Pb correlations persist even after the local octahedral tilt correlations are removed: in the orthorhombic $Pnma$ phase, the rod-like diffuse scattering from tilts is replaced by Bragg peaks, yet the flower-like pattern remains (\cref{fig:Fig2}f).

~

To resolve the origin of these two signatures, we computed the energy-resolved $S(\mathbf{q},\omega)$ from MD simulations of \ce{CsPbBr3} (\cref{fig:Fig2}d).
Integrating only the quasi-elastic window ($E<1$~meV) reproduces the $R$- and $M$-point rods but not the flower pattern (\cref{fig:Fig2}d, left), confirming that the rods are QEDS from the soft overdamped tilt modes, as established previously \cite{dubajic_dynamic_2025,lanigan-atkins_two-dimensional_2021, fransson_limits_2023}.
Integrating instead the 2--3~meV window recovers the flower pattern while the rods disappear (\cref{fig:Fig2}d, top right), demonstrating that the flower originates from regular underdamped phonons rather than from overdamped modes (\suppfigref{S-fig:CsPbBr3_Energy_Windows}).
This distinguishes the two signatures: the $R$- and $M$-point rods are quasi-elastic, originating from soft overdamped phonons with frequencies approaching zero (as in the displacive phase-transition limit) that give rise to the dynamic tilt nanodomains, whereas the flower pattern is inelastic, arising from phonons that do not exhibit soft-mode behaviour.
Decomposing $S(\mathbf{q},\omega)$ into atomic partials shows that the flower pattern is reproduced by the sum of the Pb--Pb and Cs--Pb contributions (\cref{fig:Fig2}c, bottom right), confirming that it arises from correlated Pb--Pb and Cs--Pb displacements, consistent with the \ce{SrTiO3} analogue \cite{kopecky_diffuse_2012}.
The same correlation is present in the MA- and FA-based perovskites (\suppfigref{S-fig:APb_flower_FA_MA}) but leaves a weaker fingerprint in total X-ray scattering because the molecular A-site cation has a much smaller form factor than Cs, suppressing the cross-channel contribution (\suppsecref{sup_ssec:CrossMaterial_visibility}).

~

These observations directly address a long-standing debate in halide perovskites: whether the B-site (Pb) cation undergoes local off-centering, as inferred from pair distribution function studies of the cubic phase \cite{laurita_chemical_2017}, or whether these displacement signatures are purely thermal in origin.
The flower-like scattering shows that cation-displacement correlations, with finite correlation lengths, are indeed present.
A single phonon mode is delocalised, with effectively infinite spatial coherence; the finite correlation length emerges from the thermal population and dispersion of many such delocalised modes, exactly as in the conventional thermal diffuse scattering obtained from a harmonic phonon dispersion.
Because the Cs and Pb correlations arise from regular underdamped phonons and not overdamped soft modes, their displacement amplitudes, which scale inversely with phonon frequency squared, are small compared with those of the soft octahedral tilt modes.
We therefore conclude that these cation-displacement correlations are real and measurable, but that their dominant contribution is conventional phonon thermal diffuse scattering.
Within the quasi-elastic channel probed here, we find no evidence for a soft, symmetry-defining local off-centering, and the dynamic octahedral tilt nanodomains remain the dominant local structure in LHPs.
\cref{fig:Fig2}f illustrates the evolution of $S(\mathbf{q})$ through the phase transitions in \ce{CsPbBr3}, with the $R$ and $M$ points indicated.
The cubic-to-tetragonal transition between \SI{400}{\kelvin} and \SI{395}{\kelvin} is signified by the emergence of Bragg peaks at the $M$ point, and the tetragonal-to-orthorhombic transition between \SI{365}{\kelvin} and \SI{355}{\kelvin} is signified by the additional appearance of Bragg peaks at the $R$ and $X$ points.
Overall, these results underscore the central role of the A-site cation in controlling local structural correlations in LHPs.

\subsection*{Halide engineering of local structure in mixed halide LHPs}

\begin{figure}[H]
    \centering
    \includegraphics[width=0.75\textwidth]{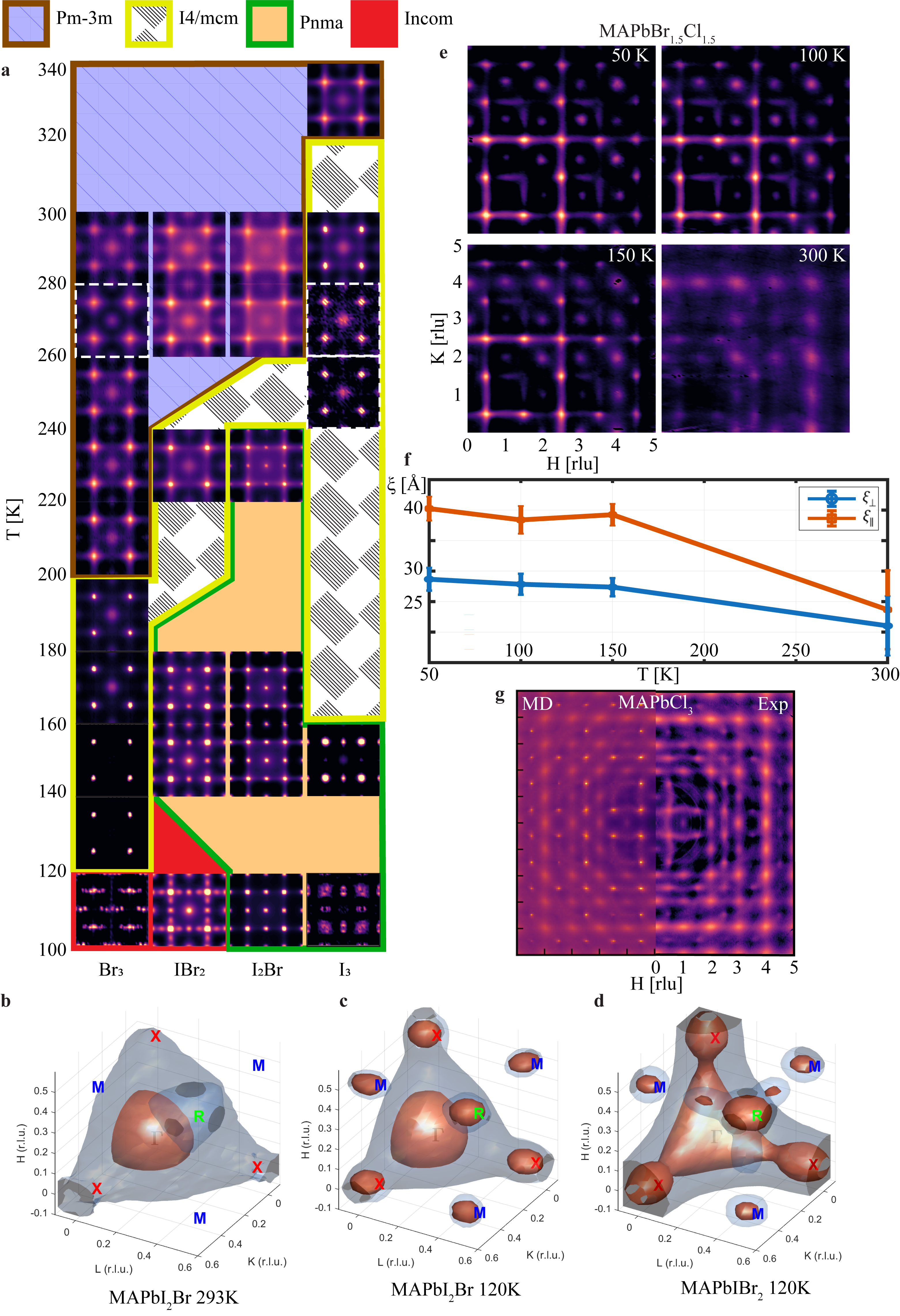}
\caption{{\bf Halide substitution is an efficient way to engineer local structure in mixed halide LHPs.}
\textbf{a,}~Phase diagram of the \ce{MAPb(Br_{x}I_{3-x})} as a function of temperature and composition.
Each square shows a cross-section of the folded first BZ at $l = 0.5$, containing the $R$, $M$, and $X$ high-symmetry points; $S(\mathbf{q})$ is plotted on a logarithmic colour scale with the same intensity range for every composition, allowing direct comparison of the gradual increase in diffuse scattering intensity as the phase transition is approached.
Coloured regions indicate the cubic ($Pm\bar{3}m$), tetragonal ($I4/mcm$), orthorhombic ($Pnma$), and incommensurate phases.
Squares with white dashed outlines show the folded BZs obtained from MD simulations. Empty areas correspond to unmeasured temperature ranges.
\textbf{b--d,}~3D isosurface representations of the folded first BZ for representative compositions and temperatures.
The opaque isosurface corresponds to a high isovalue and highlights Bragg peaks of the average structure, while the transparent isosurface corresponds to a lower isovalue revealing diffuse scattering.
\textbf{e,}~$S(\mathbf{q})$ in the $\text{HK}1.5$ plane (logarithmic colour scale) for \ce{MAPbBr_{1.5}Cl_{1.5}} at \SI{50}{\kelvin}, \SI{100}{\kelvin}, \SI{150}{\kelvin}, and \SI{300}{\kelvin}, showing progressive sharpening and intensification of the $R$-point diffuse scattering upon cooling in the absence of a phase transition.
\textbf{f,}~In-plane ($\xi_{\parallel}$) and out-of-plane ($\xi_{\perp}$) correlation lengths extracted from the diffuse scattering in panel \textbf{e} as a function of temperature. Error bars represent 95\% confidence intervals of the performed fits. 
\textbf{g,}~$S(\mathbf{q})$ in the $\text{HK}1.5$ plane (logarithmic colour scale) for \ce{MAPbCl3} at \SI{200}{\kelvin}: MD simulation (left) and experiment (right). Both show diffuse scattering rods with a distinctive curvature, suggesting additional correlations beyond simple octahedral tilting.}
    \label{fig:Fig3}
\end{figure}

To investigate how X-site composition controls the behaviour of the local and average structure, we map the \ce{MAPb(Br_{x}I_{3-x})} series across temperature and composition using single crystal X-ray diffuse scattering.
\cref{fig:Fig3}a presents the resulting phase diagram, constructed by folding the full three-dimensional (3D) reciprocal-space volume into the first Brillouin zone (BZ) (\suppsecref{sup_sec:BZ_Folding}).
Each square in the phase diagram corresponds to a 2D hk cross-section of the folded BZ at $l = 0.5$, which contains the $R$, $M$, and $X$ high-symmetry points; the nature of intensity features at these points directly reveals the local or long-range structural order present.
Squares with white dashed outlines in \cref{fig:Fig3}a show the corresponding folded BZs from MD simulations, which closely reproduce the experimental patterns and confirm that the simulations capture the overall trends in local structure across the compositional series.
Three phases are identified across the compositional series: cubic ($Pm\bar{3}m$), tetragonal ($I4/mcm$), and orthorhombic ($Pnma$), with an additional incommensurate region at the lowest temperatures for the bromide-rich \ce{MAPbBr3} and \ce{MAPbIBr2}.

~

In the cubic phases, diffuse scattering at the $R$ points is consistently observed across all compositions (\cref{fig:Fig3}b), confirming the presence of local anti-phase tetragonal ($a^{0}a^{0}c^{-}$) nanodomains with pancake-shaped morphology.
As the temperature is lowered, the diffuse peaks gradually sharpen and their intensity increases, reflecting an increase in both correlation length and dynamic disorder (as also confirmed from MD, \suppsecref{sup_sec:SigmaTheta_vs_T}), until they eventually condense into the sharp Bragg reflections of the average tetragonal or orthorhombic phases.
The cubic-to-tetragonal phase transition temperature varies non-monotonically with composition: the \ce{I2Br} and \ce{IBr2} compositions exhibit higher transition temperatures than either the neat bromide or neat iodide end members.
Across all compositions except in \ce{MAPbBr3}, the orthorhombic $Pnma$ phase is reached at sufficiently low temperatures (see phase diagram in \cref{fig:Fig3}a to track transition temperatures).
3D BZ volumes for representative compositions are shown in \cref{fig:Fig3}b-d.
In cubic \ce{MAPbI2Br} at \SI{293}{\kelvin} (\cref{fig:Fig3}b), the $\Gamma$ Bragg peaks of the cubic structure are accompanied by diffuse intensity at the $R$ point, confirming local tetragonal character.
In orthorhombic \ce{MAPbI2Br} at \SI{120}{\kelvin} (\cref{fig:Fig3}c), the $R$, $M$, and $X$ reflections all appear as Bragg peaks, as expected for $Pnma$ symmetry.
In \ce{MAPbIBr2} at \SI{120}{\kelvin} (\cref{fig:Fig3}d), an incommensurate structure is observed: the same $Pnma$ Bragg peaks are present, but three additional satellite reflections appear around the $R$ point, which we assign to incommensurate modulation of the octahedral tilts.
Temperature-dependent $\text{HK}1.5$, $\text{HK}0.5$, and $\text{HK}3.5$ reciprocal space planes for the \ce{I2Br} and \ce{IBr2} compositions are provided in \suppfigref{S-fig:HK15_Compositional_Series} and full temperature-dependent $\text{HK}1.5$ data for \ce{MAPbBr3}, \ce{MAPbI3}, \ce{MAPbCl3}, and \ce{MAPbBr_{1.5}Cl_{1.5}} are shown in \suppfigref{S-fig:HK15_Pure_Halides}.

~

Our previous study of the \ce{MAPb(Br_{x}I_{3-x})} series using powder X-ray diffraction reported the coexistence of multiple crystallographic phases over a wide temperature range and identified a monoclinic distortion \cite{shahrokhi_anomalous_2022}.
The present single crystal diffraction data show that QEDS from local nanodomains, when averaged over a powder sample, mimics the superstructure reflections of additional long-range-ordered phases, giving a false impression of phase coexistence.
We thus conclude that in these mixed halide compositions a single average phase coexists with short-range ordered nanodomains at each temperature, rather than a mixture of distinct long-range ordered phases.

~

To investigate the temperature evolution of the local structure in the absence of competing phase transitions, we focus on the mixed bromide-chloride composition \ce{MAPbBr_{1.5}Cl_{1.5}} (\cref{fig:Fig3}e), which remains in the cubic phase down to the lowest temperature we measured (\SI{50}{\kelvin}) \cite{BARI_crystal}.
This composition effectively contains a quenched local structure: as the temperature decreases from \SI{300}{\kelvin} to \SI{50}{\kelvin}, the diffuse scattering at the $R$ point progressively sharpens and intensifies, indicating that both the correlation lengths and the dynamic disorder increase upon cooling (as also confirmed from MD, \suppsecref{sup_sec:SigmaTheta_vs_T}).
The extracted in-plane ($\xi_{\parallel}$) and out-of-plane ($\xi_{\perp}$) correlation lengths are plotted as a function of temperature in \cref{fig:Fig3}f, confirming that the size of the dynamic nanodomains increases.
Based on our previous findings \cite{dubajic_dynamic_2025}, increased correlation lengths are accompanied by longer nanodomain lifetimes. These findings indicate that the local structure in \ce{MAPbBr_{1.5}Cl_{1.5}} evolves continuously from a weakly correlated, short-lived state at room temperature towards a strongly correlated, longer-lived state at low temperatures, without ever undergoing a phase transition.
Finally, we examine \ce{MAPbCl3} (\cref{fig:Fig3}g).
At \SI{200}{\kelvin}, approximately \SI{50}{\kelvin} above the cubic-to-tetragonal phase transition, diffuse scattering rods are observed with a distinctive curvature in reciprocal space, which our MD simulations reproduce (\cref{fig:Fig3}g, left).
This curvature likely indicates the presence of additional correlations beyond simple octahedral tilting (\suppsecref{sup_sec:MAPbCl3_Curvature}).
At room temperature, the QEDS is nearly absent (\suppfigref{S-fig:MAPbCl3_HK15_TempSeries}), suggesting that the local structure in \ce{MAPbCl3} is only weakly developed and appears primarily upon approaching the phase transition.
MD simulations with different halide mixing rule out halide X-site correlations as the origin of these diffuse scattering features (\suppfigref{S-fig:HalideDistribution_CoolingRate}).

~

Together, these results demonstrate that halide composition provides a powerful lever for engineering the local structural landscape: by tuning the X-site, it is possible to shift phase transition temperatures, control the dynamic disorder and the correlation length of nanodomains, quench the local structure into a single phase without a competing transition, and introduce incommensurate modulations, all of which have the potential to influence macroscopic optoelectronic properties.
\subsection*{Thermal-history control of local structure, phase transitions and optoelectronic response}

While the optoelectronic properties of LHPs are widely studied as a function of static temperature and composition, the role of the thermal history during processing and operation remains a critical yet underexplored variable. 
Understanding how heating and cooling rates influence structural evolution is essential, as phase transitions in these soft lattices may be tied to device performance and long-term stability. 

~

Previous studies have reported considerable ambiguity in the low-temperature phase sequence of \ce{FAPbBr3}: while a consensus exists on the space groups of each phase, the transition temperatures and the precise sequence of intermediate phases have remained unclear \cite{sharma_contrasting_2020,govinda_critical_2018,mozur_dynamical_2019,schueller_crystal_2018,keshavarz_tracking_2019,franz_influence_2020}.
Our results suggest that this ambiguity may originate from the sensitivity of \ce{FAPbBr3} to cooling rates, combined with the presence of incommensurate structures that depend on thermal history.
As shown in \cref{fig:Fig4}a, quenching (left panel) and fast cooling (top right panel) both induce incommensurate structures, characterised by satellite reflections around the $R$ and $M$ points.
The incommensurate modulation differs depending on the cooling rate: different satellite peak positions are observed for quenched versus fast-cooled crystals, indicating that the cooling rate tunes the modulation wavevector.
By contrast, slow cooling (bottom right panel) yields the conventional $a^{+}a^{+}a^{+}$ ($Im\bar{3}$) phase without any incommensurate structure.
We also confirmed the incommensurate structure under fast cooling using single crystal neutron diffraction (\suppfigref{S-fig:FAPbBr3_Neutron}).

~

We also investigated \ce{FAPbI3} by differential scanning calorimetry (DSC) (\cref{fig:Fig4}c).
At \SI{3}{\kelvin/\minute} the cubic ($a^{0}a^{0}a^{0}$) to cubic ($a^{+}a^{+}a^{+}$) phase transition appears as a pronounced peak, reproduced across samples, whereas at \SI{0.05}{\kelvin/\minute} the peak is absent indicating that the transition may be suppressed. The peak is likewise absent at every intermediate rate we measured, so the transition is only resolved on cooling between 1.6 and \SI{3}{\kelvin/\minute} (\suppsecref{sup_sec:FAPbI3_CoolingRate}). These results are consistent with the cooling-rate-dependent single crystal X-ray diffraction patterns reported for \ce{FAPbI3} \cite{liang_fapi_twin}.
Molecular dynamics simulations provide a real-space picture of how thermal history shapes the \ce{FAPbI3} structure \cite{liang_fapi_twin}. In its ordered phase, \ce{FAPbI3} develops long-range, coherent $a^{+}a^{+}a^{+}$ in-phase tilts (\cref{fig:Fig4}d). On cooling, however, the FA molecular reorientation freezes out before the octahedral-tilt network can order, so the system can become kinetically arrested in a metastable twin-domain state, in which locally $a^{+}a^{+}a^{+}$ nanodomains are separated by sharp twin-like boundaries (\cref{fig:Fig4}e), rather than reaching the totally ordered phase. Another ML-driven MD study found kinetic trapping in a different metastable low-temperature tilt configuration, also highlighting the role of FA reorientation/freezing and the shallow competing tilt landscape in \ce{FAPbI3} \cite{julia_fapi}.
These findings together demonstrate that the average and local structure in \ce{FAPbI3} are strongly dependent on thermal history. 

~

In \ce{MAPbI3} (\cref{fig:Fig4}b), we compare single crystal diffraction patterns for two cooling rates: \SI{2}{\kelvin/\minute} and \SI{0.02}{\kelvin/\minute}.
At the conventional cooling rate of \SI{2}{\kelvin/\minute}, the cubic-to-tetragonal phase transition produces the expected $a^{0}a^{0}c^{-}$ ($I4/mcm$) tetragonal phase, with the same local structure ($a^{0}a^{0}c^{-}$) as shown in \cref{fig:Fig1}a,b and \suppfigref{S-fig:Schematic_Case5}.
However, at the much slower rate of \SI{0.02}{\kelvin/\minute}, the system instead adopts the $a^{0}a^{0}c^{+}$ ($P4/mbm$) tetragonal phase, with an underlying local orthorhombic ($a^{-}a^{-}c^{+}$) structure reminiscent of the Cs-based systems (\cref{fig:Fig1}a, left; \suppfigref{S-fig:Schematic_Case3}; both tilt systems assigned in \suppsecref{sup_sec:a0a0cplus_MAPbI3}).
This observation of $a^{0}a^{0}c^{+}$ in \ce{MAPbI3} is unusual, as the conventional tetragonal phase is universally reported as $a^{0}a^{0}c^{-}$, although in-phase tilting has been predicted as a competing metastable state in computational studies \cite{fransson_revealing_2023}. To rule out X-ray beam damage as the origin, we repeated the measurement at constant temperature multiple times and observed no change in the diffraction pattern (\suppsecref{sup_sec:BeamDamage_MAPbI3}).
\begin{figure}[H]
    \centering
    \includegraphics[width=1\textwidth]{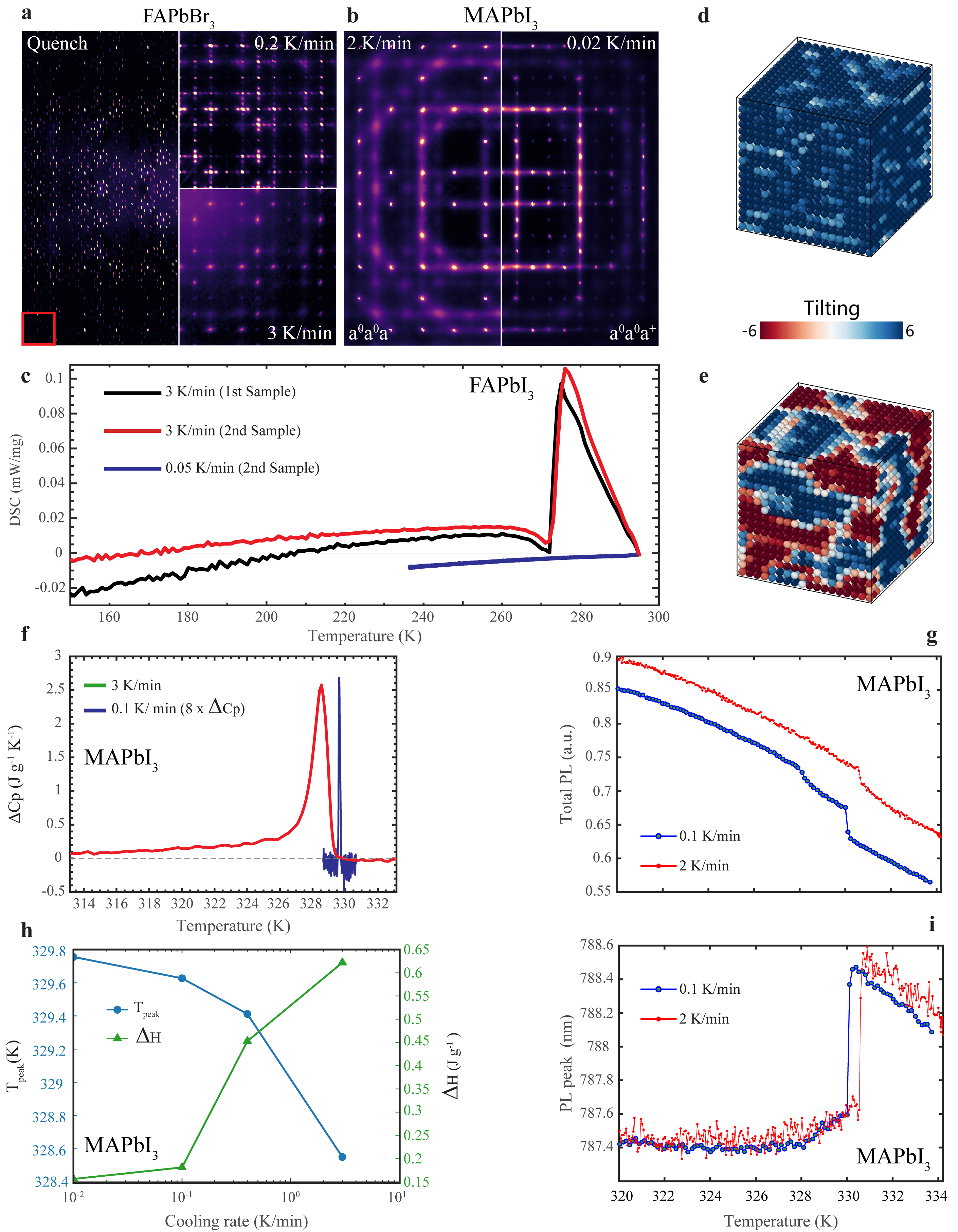}

\end{figure}

\afterpage{%
\begin{figure}[t]
\caption{{\bf Thermal history dictates phase, local structure and optoelectronic properties in LHPs.}
\textbf{a,}~$S(\mathbf{q})$ in the $\text{HK}1.5$ plane for \ce{FAPbBr3} under three cooling conditions: quenched to 102 K (left), fast-cooled to 100 K (top right), and slow-cooled to 130 K (bottom right). The red square in the bottom-left corner serves as a scale bar, representing $1 \times 1\text{ r.l.u.}$ along the respective reciprocal-space dimensions.
\textbf{b,}~$S(\mathbf{q})$ in the $\text{HK}1.5$ plane for \ce{MAPbI3} at two cooling rates. At \SI{2}{\kelvin/\minute} (left), the conventional $a^{0}a^{0}c^{-}$ ($I4/mcm$) tetragonal phase is observed at 300K. At \SI{0.02}{\kelvin/\minute} (right), the unconventional $a^{0}a^{0}c^{+}$ ($P4/mbm$) tetragonal phase with local orthorhombic ($a^{-}a^{-}c^{+}$) structure is observed at 335K.
\textbf{c,}~Differential scanning calorimetry (DSC) of \ce{FAPbI3}: the phase-transition peak is pronounced at \SI{3}{\kelvin/\minute} (two samples) but absent at \SI{0.05}{\kelvin/\minute}.
\textbf{d,e,}~Real-space visualisation of the \ce{FAPbI3} octahedral tilts from MD simulations \cite{liang_fapi_twin}, with each octahedron represented by a sphere coloured by its tilt angle along the three Cartesian directions: the ordered $a^{+}a^{+}a^{+}$ phase (\textbf{d}) and the kinetically arrested network of static $a^{+}a^{+}a^{+}$ nanodomains separated by twin-like boundaries (\textbf{e}). Colorbar values are given in degrees. 
\textbf{f,}~Excess heat capacity $\Delta C_p$ of \ce{MAPbI3} at \SI{3}{\kelvin/\minute} and \SI{0.1}{\kelvin/\minute} (the \SI{0.1}{\kelvin/\minute} trace scaled $\times 8$ for visibility), showing a pronounced tail at the faster rate and a $\sim$\SI{1}{\kelvin} shift of the transition.
\textbf{g,}~Temperature-dependent integrated photoluminescence (PL) of \ce{MAPbI3} at two heating rates; the dip at the transition is deeper for slower heating.
\textbf{h,}~Transition peak temperature $T_\mathrm{peak}$ and released enthalpy $\Delta H$ of \ce{MAPbI3} as a function of cooling rate.
\textbf{i,}~PL peak position of \ce{MAPbI3} versus temperature at two heating rates.}
\label{fig:Fig4}
\end{figure}%
}
DSC measurements on \ce{MAPbI3} show that this cooling-rate sensitivity is imprinted on the cubic-to-tetragonal transition itself. The excess heat capacity $\Delta C_p$, which isolates the latent heat, forms a sharp peak at \SI{0.1}{\kelvin/\minute} but broadens markedly and develops an extended low-temperature tail at \SI{3}{\kelvin/\minute}, with the peak shifting $\sim$\SI{1}{\kelvin} lower in temperature at the faster rate (\cref{fig:Fig4}f). Across cooling rates the peak temperature $T_\mathrm{peak}$ decreases and the integrated enthalpy $\Delta H$ grows several-fold (\cref{fig:Fig4}h). This observation of kinetic supercooling, together with the non-conserved latent heat released over a long tail below the transition, is the signature of a ferroelastic transition whose order-parameter relaxation cannot keep pace with the imposed ramp, driving it progressively out of equilibrium at faster cooling (\suppsecref{sup_sec:DSC_MAPbI3}).

~

These structural and thermodynamic changes translate directly into the optoelectronic response. Temperature-dependent photoluminescence (PL) on \ce{MAPbI3}, measured on heating, shows a $\sim$\SI{1}{\kelvin} heating-rate-dependent shift in the transition (\cref{fig:Fig4}g,i). The dip in integrated PL intensity at the transition is substantially deeper for slower heating, so that faster heating preserves a higher PL intensity (\cref{fig:Fig4}g), while the PL peak energy shifts slightly, with faster heating yielding a marginally redder emission (\cref{fig:Fig4}i; underlying spectra and the definitions of both quantities in \suppsecref{sup_sec:PL_MAPbI3}). Because PL intensity directly scales with the photoluminescence quantum yield (PLQY), which, according to standard semiconductor equations, dictates the open-circuit voltage ($V_{\text{oc}}$),these results imply that the $V_{\text{oc}}$ of perovskite solar cells will be highly sensitive to heating and cooling rates.  Thermal history thus provides a direct handle on the optoelectronic response of a fixed composition, demonstrated here for \ce{MAPbI3}.

~

\subsection*{Conclusions}

By mapping the local dynamic structural landscape of LHPs across a broad compositional and temperature space, we have established that dynamic octahedral tilt nanodomains are a universal feature of the LHPs: almost every composition studied hosts equilibrium local structural fluctuations in part of its phase diagram. Their character is set by three complementary levers. The A-site cation sets the nanodomain symmetry and anisotropy, with local disorder increasing from FA through MA to Cs; the halide sets the dynamic disorder, which increases from chloride through bromide to iodide; and thermal history selects which average phase, and which local order within it, the material adopts. 

~

The nature of this local disorder directly tracks the optoelectronic quality of the corresponding materials.
We previously established this link in two bromide compositions: the denser, more anisotropic nanodomains of \ce{MAPbBr3}, compared to \ce{FAPbBr3}, coincide with greater electronic disorder (larger Urbach energies) and lower photoluminescence quantum yield (PLQY) \cite{dubajic_dynamic_2025}.
Here we show that the weakly disordered, isotropic nanodomains of \ce{FAPbBr3} are not specific to that composition, but are common to every FA-based system studied, including \ce{FAPbI3} and the FA-dominated triple cation alloy.
FA-based perovskites therefore sit as a family at the least disordered end of this structural landscape, consistent with the superior performance of FA-rich absorbers across optoelectronic applications.
The trend we observe, with local disorder increasing from FA through MA to Cs, is reproduced independently by first-principles calculations of polymorphous LHPs, where the electron-phonon coupling contribution to band gap renormalisation increases along the same series in both the bromides and the iodides \cite{zacharias_electron-phonon_2026,zacharias_roadmap_2026}.
Empirically, thin film devices reach 90\% of the detailed balance limit for iodide-rich compositions, around 73\% for bromide and mixed I/Br compositions, and roughly 10\% for chloride-only compositions \cite{PVtables_v6,cl_efficiency}.
This ordering tracks fundamental properties known to vary with halide substitution: defect tolerance \cite{meggiolaro_iodine_2018}, Fröhlich coupling to longitudinal optical phonons \cite{herz_charge-carrier_2017,kang_mobility_2018}, and non-radiative recombination \cite{zhang_radiative_2018,zhang_carrier_2020}.
We find that the dynamic disorder in single crystals increases from chloride through bromide to iodide, and propose that this local structural trend is the more fundamental origin underlying these properties.

~

Experimentally, we find that thermal history alone determines both the average phase and the local structure within it. Across \ce{FAPbBr3}, \ce{MAPbI3} and \ce{FAPbI3}, the cooling rate drives nominally identical compositions into different average and local phases, incommensurate modulations, or kinetically arrested metastable states. These states are not optoelectronically equivalent: in \ce{MAPbI3} the heating rate alone changes the PLQY through the transition implying that the $V_{\text{oc}}$ of a fixed composition may be strongly thermal-history dependent. We demonstrate this coupling for \ce{MAPbI3}, but as thermal history governs the structure of multiple compositions studied here, we expect this finding to be universal across LHPs.

~

The range of cooling rates used in our experiments, \SI{0.02}{\kelvin/\minute} to \SI{3}{\kelvin/\minute}, lies within that of IEC 61215 stability testing, which involves thermal cycling at rates up to $1.66~\text{K/min}$ and humidity-freeze tests up to $3.33~\text{K/min}$ \cite{holzhey_stability_2018}, with the effective ramp rate set by the thermal mass of the module. The phase transitions of LHPs lie within the operating window of devices: terrestrial protocols cycle between $-40^\circ\text{C}$ and $85^\circ\text{C}$ for 200 cycles, and space missions impose temperatures from $-180^\circ\text{C}$ to $120^\circ\text{C}$ over as many as 6,500 cycles \cite{yilmaz_thermal_2026}. Devices are therefore driven through the same transitions, at the same rates, that we show here to determine the local structure and the optoelectronic response. The crystallographic structure and optoelectronic response of a device thus may not only be fixed by its composition but can be re-established by every thermal cycle it undergoes. Because this proceeds without any chemical change, it is distinct from ionic motion, halide segregation and electrochemical degradation, that may also be independently affecting the performance of devices during these tests.
~

~

Our results reveal that the structural landscape of LHPs is far broader than conventionally assumed.
Even nominally identical compositions can adopt multiple distinct local and average structures depending on their thermal history, so their optoelectronic properties can be modulated beyond what composition alone allows.
This flexibility presents both a challenge, in that structural characterisation must account for hidden local order and thermal history, and an opportunity, in that nanodomains can be deliberately engineered through composition, synthesis and thermal history to optimise the optoelectronic response for photovoltaics, light emission and detection.

\section*{Methods}
\label{sec: Methods}

\subsection*{Synthesis of perovskite single crystals}

\ce{MAPbBr_{3}}: \ce{MAPbBr3} single crystals were provided by AY Sensors Inc.\ (\url{https://www.aysensors.com/}, Canada), grown by flux-regulated crystallization \cite{haruta_reproducible_2024}. Such crystals can also be grown by inverse temperature crystallisation \cite{saidaminov_high-quality_2015}: A mixture of 1M of PbBr$_2$ and 1M MABr was dissolved in 1 mL DMF. To ensure complete dissolution, the solution was stirred vigorously at 25 $^{\circ}$C for 6 h and then filtered with a $\SI{0.45}{\micro\meter}$ filter head before use.
After adding a small MAPbBr$_3$ crystal to the filtered solution, the solution was transferred to an oven. To make the crystals larger, the solution was further heated to 85 $^{\circ}$C with a rate of 10 $^{\circ}$C per 30 min, and the crystal reached its full size after 24 hours. The obtained crystals were separated and dried to obtain MAPbBr$_3$.

~

\ce{FAPbBr_{3}}: \ce{FAPbBr3} single crystals were grown by inverse temperature crystallisation \cite{saidaminov_retrograde_2015}. A mixture of 1M of PbBr\textsubscript{2} and 1 M FABr was dissolved in 1 mL DMF/GBL (1:1). To ensure complete dissolution, the solution was stirred vigorously at 25 $^{\circ}$C for 6 h and then filtered with a $\SI{0.45}{\micro\meter}$ filter head before use. After adding a small FAPbBr$_3$ crystal to the filtered solution, the solution was transferred to an oven. To make the crystals larger, the solution was further heated to $55\degree$ C with a rate of 10 $^{\circ}$C per 30 min, and the crystal reached its full size after 24 hours. The obtained crystals were separated and dried to obtain FAPbBr$_3$.

~

\ce{FAPbI3} (single crystals): \ce{FAPbI3} single crystals were grown in \ce{N2} from seed crystals following Ref.~\cite{duijnstee_understanding_2023}. Equimolar FAI and \ce{PbI2}, together with 3.8 mol\% \ce{MDACl2}, were dissolved in GBL. The solution was stirred at 60 $^{\circ}$C for four hours, then filtered with a 25 mm diameter $\SI{0.45}{\micro\meter}$ GMF filter. Seed crystals were grown from this solution by rapid heating to 140 $^{\circ}$C, collected, and placed in an empty vial, to which 4 mL of the same precursor solution was added. The vial was kept undisturbed in an oil bath at 95 $^{\circ}$C for 12 hours until black crystals of 1--4 mm size appeared. The crystals were then dried in a vacuum oven at 180 $^{\circ}$C for 45 minutes.

~

\ce{FAPbI3} (powder): \ce{PbI2} (6.5 mmol) and FAI (7.1 mmol, 10\% excess) were dissolved in anhydrous 2ME (3.65 mL) with rapid stirring over two hours to obtain a clear yellow solution. The solution was filtered through a $\SI{0.22}{\micro\meter}$ PTFE pad into a vial and immersed in a preheated oil bath at 120 $^{\circ}$C for 45 minutes with stirring, precipitating the black phase. The hot suspension was vacuum filtered and the precipitate rinsed thoroughly with anhydrous MeCN. The powder was annealed in air at 150 $^{\circ}$C for 30 minutes, dried under vacuum overnight at ambient temperature, then stored under nitrogen in amber glass (yield: 85\%).

~

\ce{MAPbI3}: \ce{MAPbI3} single crystals were grown by inverse temperature crystallisation \cite{saidaminov_high-quality_2015}. A mixture of 1 M \ce{PbI2} and 1 M MAI was dissolved in 1 mL GBL. The solution was heated from room temperature and kept at 115 $^{\circ}$C to initiate crystallisation. The crystal reached its full size after 3 hours.

~

\ce{MAPbI2Br} and \ce{MAPbIBr2}: The mixed-halide single crystals were grown by inverse temperature crystallisation \cite{saidaminov_high-quality_2015}. Methylammonium and lead halide precursors (\ce{MAI}, \ce{MABr}, \ce{PbI2} and \ce{PbBr2}) were combined in the target halide ratio and dissolved to 0.9 M in mixed GBL/DMF solvent, using GBL:DMF volume ratios of 2:1 for \ce{MAPbI2Br} and 1:2 for \ce{MAPbIBr2}. Each solution was stirred at 60 $^{\circ}$C until fully dissolved and filtered through a $\SI{0.45}{\micro\meter}$ PTFE filter. A 3 mL aliquot of the filtrate was then heated on a hotplate, with the temperature ramped to 120 $^{\circ}$C to induce crystallisation, and the crystals were grown for 3 hours under ambient conditions.

~

\ce{MAPbCl3}: \ce{MACl} and \ce{PbCl2} were dissolved at a 1:1 molar ratio at a Pb concentration of 1 M in a DMSO/DMF mixed solvent with a volume ratio of 1:1. The precursor solution was then filtered through a $\SI{0.22}{\micro\meter}$ PTFE filter. The filtered solution was transferred into a sealed vial, which was immersed in a water bath at 50 $^{\circ}$C to promote single-crystal growth. After seed single crystals had formed, they were collected and transferred into a fresh saturated \ce{MAPbCl3} precursor solution for further growth.

~

\ce{MAPbBr_{1.5}Cl_{1.5}}: A \ce{MAPbBr_{1.5}Cl_{1.5}} precursor solution was prepared by dissolving 0.8 M \ce{MABr}, 0.8 M \ce{MACl}, 0.64 M \ce{PbBr2}, and 0.64 M \ce{PbCl2} in 2 mL DMF and 2 mL DMSO at 50 $^{\circ}$C with overnight stirring. The solution was filtered with a PTFE filter ($\SI{0.45}{\micro\meter}$ pore size). A small glass vial was filled with 4 mL of the filtered solution and sealed with foil containing small pinholes to allow dichloromethane (DCM) vapour to slowly dissolve into the perovskite precursor solution. DCM was used as an anti-solvent to precipitate the single crystals. The glass vial was stored under a DCM atmosphere, and the \ce{MAPbBr_{1.5}Cl_{1.5}} single crystals grew continuously for two days. All the procedures were performed under ambient conditions.

~

\ce{CsPbBr3} and \ce{CsPbCl3} (Bridgman growth): \ce{CsPbBr3} and \ce{CsPbCl3} single crystals were grown by the Bridgman melt-growth method. Binary precursor salts were synthesised from \ce{Pb(CH3COO)2.3H2O} (Thermo Scientific, $>$99\%), \ce{Cs2CO3} (99.999\%), \ce{HBr} (48 wt\% in \ce{H2O}, Thermo Scientific), and \ce{HCl} ($\geq$37 wt\% in \ce{H2O}, Honeywell ACS reagent). Stoichiometric mixtures of \ce{CsBr} and \ce{PbBr2}, or \ce{CsCl} and \ce{PbCl2}, were reacted at 650 $^{\circ}$C to produce polycrystalline \ce{CsPbBr3} and \ce{CsPbCl3}, respectively. The resulting material ($<$30 g) was loaded into 16:18 mm fused-silica ampoules and sealed under vacuum ($\sim 10^{-4}$~Torr) for Bridgman growth, after manual removal of visible dark impurity residues from the initial precursors \cite{chung_growth_2024}. High-quality single crystals were grown by sequential directional solidification using the vertical Bridgman technique, with translation rates of 3 mm~h$^{-1}$ and 1 mm~h$^{-1}$. For \ce{CsPbBr3}, the precursor was heated to 590 $^{\circ}$C, held for 8 h, and homogenised at the maximum temperature for a further 10 h prior to growth; for \ce{CsPbCl3}, the precursor was heated to 650 $^{\circ}$C. \ce{CsPbCl3} crystals were grown in a two-zone vertical Bridgman furnace, whereas \ce{CsPbBr3} crystals were grown in a three-zone Bridgman furnace using a temperature profile adapted from previous studies \cite{he_demonstration_2021,he_high_2018}. During growth, impurities preferentially segregated to the top and bottom of the ingot and were removed after the fast-growth stage (3 mm~h$^{-1}$), leaving a central single-crystalline portion; the final growth was carried out at 1 mm~h$^{-1}$.

~

\ce{CsPbCl3} (solution growth): Solution-grown \ce{CsPbCl3} crystals were additionally prepared. To a 20 mL scintillation vial were added 565.5 mg \ce{CsCl} (50\% molar excess, Sigma-Aldrich, 99.9\%), 499.9 mg \ce{PbO} (Alfa Aesar, 99.9\%), and 10 mL \ce{HCl} (Sigma-Aldrich, 37\% w/w). The mixture was stirred at 110 $^{\circ}$C for 30 min, after which stirring was halted and the heated mixture was left for 30 min to allow the remaining solids to settle. The clear solution was decanted into a new vial, leaving the undissolved solids behind, heated at 130 $^{\circ}$C for one hour, and then slowly cooled to room temperature over 15 h, producing small yellow crystals of \ce{CsPbCl3}. The crystals were washed with diethyl ether under vacuum and stored in a dry, dark environment until shipment.

~

Triple cation \ce{Cs_{0.05}(MA_{0.17}FA_{0.83})_{0.95}Pb(I_{0.83}Br_{0.17})3}: The triple-cation mixed-halide precursor solution was prepared by co-dissolving \ce{PbI2} (1.1 M), \ce{PbBr2} (0.2 M), FAI (1.1 M), MABr (0.13 M) and \ce{CsBr} (0.065 M) in 50 mL of GBL. The mixture was stirred on a hotplate at approximately 50--60 $^{\circ}$C until a clear, transparent, particle-free solution was obtained. Formic acid (FAH) was then added at 0.8 vol\% relative to the solvent (0.4 mL FAH per 50 mL GBL), and the solution was stirred for a further $\sim$30 min to ensure homogeneity. Prior to crystal growth, the solution was filtered through a $\SI{0.2}{\micro\meter}$ PTFE syringe filter to remove undissolved particulates and heterogeneous nucleation sites. The filtered precursor solution was transferred into clean glass vials, which were slowly heated from 60 $^{\circ}$C to 68 $^{\circ}$C at a rate of 1 $^{\circ}$C~h$^{-1}$ ($\approx$8 h total). The slow heating rate promotes controlled nucleation and growth of large, single-domain crystals. Once crystals of the desired size had formed, they were extracted from the hot solution, the residual solvent was quickly wiped from the surfaces with filter paper, and the crystals were dried in vacuum.

\subsection*{Single crystal X-ray diffuse scattering }

Single crystal X-ray diffuse scattering was measured at three beamlines of Diamond Light Source (I15, I19-2 and I19-1) and at the P21.1 beamline of PETRA~III at DESY. 

~

Measurements of \ce{CsPbBr3}, \ce{CsPbCl3}, \ce{MAPbCl3}, \ce{MAPbI3}, \ce{MAPbI2Br}, \ce{MAPbIBr2} and \ce{FAPbI3} were carried out at the I15 beamline using X-rays of \SI{72}{\kilo\eV} ($\lambda = 0.1722$~\AA). Diffraction images were recorded on a Dectris PILATUS3 X CdTe 2M detector (1679 $\times$ 1475 pixels, 172 $\times$ 172 $\mu\text{m}^2$ pixel size, \SI{1}{\milli\meter} CdTe sensor) at a sample-to-detector distance of \SI{370}{\milli\meter}, with the crystal rotated continuously about $\phi$ in \SI{0.2}{\degree} increments.

~

Measurements of \ce{CsPbBr3}, \ce{FAPbBr3} and \ce{FAPbI3} were carried out at the I19-2 beamline using X-rays of \SI{25.5}{\kilo\eV} ($\lambda = 0.4859$~\AA). Diffraction images were recorded on a Dectris EIGER2 X CdTe 4M detector (2162 $\times$ 2068 pixels, 75 $\times$ 75 $\mu\text{m}^2$ pixel size, \SI{0.75}{\milli\meter} CdTe sensor) at a sample-to-detector distance of \SI{85}{\milli\meter}, with the crystal rotated continuously about $\phi$ in \SI{0.2}{\degree} increments (fixed $\omega = \SI{-90}{\degree}$, $\kappa = \SI{0}{\degree}$).

~

Measurements of \ce{FAPbI3} at \SI{102}{\kelvin} and \ce{FAPbBr3} at \SI{100}{\kelvin} were carried out at the I19-1 beamline using X-rays of \SI{12.9}{\kilo\eV} ($\lambda = 0.9611$~\AA).
Diffraction images were recorded on a Dectris PILATUS 2M detector (1679 $\times$ 1475 pixels, 172 $\times$ 172 $\mu\text{m}^2$ pixel size, \SI{0.32}{\milli\meter} silicon sensor) at a sample-to-detector distance of \SI{160}{\milli\meter}, with the crystal rotated continuously about $\phi$ in \SI{0.2}{\degree} increments. For \ce{FAPbI3}, a thermal quench was applied by presetting the cryostream to \SI{102}{\kelvin} and inserting the room-temperature crystal directly into the stream. A lumped-capacitance estimate for a $\sim$\SI{300}{\micro\meter} crystal in a nitrogen cryostream (Biot number $\mathrm{Bi}\ll0.1$) gives an effective cooling rate during the quench of approximately 200--\SI{500}{\kelvin/\second}, roughly $10^{3}$--$10^{4}$ times faster than a standard cryostream ramp.

~

At all three Diamond beamlines the crystals were mounted on a cryoloop, coated with NVH immersion oil, and held in the nitrogen stream of an Oxford Cryostream 800 for temperature control. The three-dimensional reciprocal-space volumes were reconstructed directly from the raw detector frames in Python with the rspace3d package~\cite{rspace3d}, without an intermediate unwarping of the detector images. CrysAlisPro was used only to index the frames and to determine and refine the crystal orientation (UB) matrix. The reconstruction inverts the diffraction geometry frame by frame: for each detector pixel the scattering vector $\mathbf{q}=\mathbf{s}_1-\mathbf{s}_0$ (with $|\mathbf{s}_1|=|\mathbf{s}_0|=1/\lambda$) is built from the pixel position and the detector geometry, read from the frame header (sample-to-detector distance, pixel size, beam centre and wavelength), and is converted to a fractional Miller index $\mathbf{h}=(\mathbf{R}_{\phi}\mathbf{UB})^{-1}\mathbf{q}$, where $\mathbf{R}_{\phi}$ is the goniometer rotation of that frame. The fixed rotation that aligns the CrysAlisPro UB frame with the laboratory frame, so that Bragg reflections fall on integer positions, is recovered once by indexing a single reference frame against the known unit cell. The photon counts from every pixel of every frame are then binned onto a common cubic grid ($-6\leq h,k,l\leq 6$, voxel size $0.025$~r.l.u.), while the number of contributing pixels per voxel is accumulated separately, so that the reported intensity is the mean of the contributing measurements and voxels that receive no measurement are left undefined. Each pixel is corrected for beam polarisation and for the solid angle it subtends on the flat detector, and normalising each voxel by its number of contributing pixels supplies the Lorentz factor of the rotation geometry. Finally, hot pixels and statistical outliers are removed and the volume is averaged over the Laue symmetry of the crystal, which improves the counting statistics and fills systematically unmeasured regions, to give the final data cube.

~

X-ray diffuse scattering measurements on \ce{MAPbBr3}, \ce{FAPbBr3}, \ce{MAPbI3}, \ce{MAPbBr_{1.5}Cl_{1.5}} and the triple-cation \ce{Cs_{0.05}(MA_{0.17}FA_{0.83})_{0.95}Pb(I_{0.83}Br_{0.17})3} were conducted at the P21.1 beamline at the Positron-Elektron-Tandem-Ring-Anlage (PETRA~III) facility, Deutsches Elektronen-Synchrotron (DESY). An X-ray beam with an energy of 101.45~keV ($\lambda = 0.1222$~\AA) and a size of $0.35 \times 0.35$~mm$^2$ was used, delivering a flux of $2.5 \times 10^{10}$ photons/s. Crystals were prepared with dimensions of approximately $500 \times 500 \times 500$~$\mu$m$^3$.
Each crystal was mounted on the tip of an amorphous cactus needle using epoxy and secured to a goniometer head. The crystal was aligned to ensure it remained at the center of rotation and within the X-ray beam during measurements.
For temeprature dependent measurements N-HeliX Open Flow Helium Gas Flow System from Oxford Cryosystems was used. 
For the calibration of beam parameters Standard Reference Material® 1990, Al2O3 Single Crystal Diffractometer Alignment Standard from NIST was used. 
Diffraction images were collected using a Pilatus2M detector (1679 $\times$ 1475 pixels, 172 $\times$ 172~$\mu$m$^2$ pixel size) positioned at distances of 400~mm and 700~mm.
A single 3D dataset was obtained by rotating the crystal through a full $360^\circ$ along a single axis, with the detector continuously counting and reading out at regular intervals. Exposure time was 0.2~s per frame, producing either 1850 or 3700 total images per rotation (corresponding to $0.2^\circ$ and $0.1^\circ$ per image, respectively).
For each sample at every temperature, three different 3D datasets were collected at varying detector positions, all at the same distance, to ensure coverage of the gaps between the detector chips.
Data processing was carried out using a \textit{Matlab} repository \cite{matlab_rep}, which generated dynamic detector masks, determined crystal orientation (UB) matrices, and transformed raw detector images into reciprocal space. For diffuse scattering analysis, data were reconstructed onto a three-dimensional grid defined by $-7 \leq h,k,l \leq +7$, with voxel sizes of $\Delta h = \Delta k = 0.0274$~r.l.u. and  $\Delta l=0.0549$~r.l.u., resulting in an array of $512 \times 512 \times 256$ voxels.
After careful inspection of the diffraction data it was observed that the diffuse scattering also follows Laue symmetry. The data were subsequently averaged for Laue symmetry using a \textit{Matlab} repository \cite{matlab_rep}.

\subsection*{Single crystal neutron scattering }

Single crystal diffuse neutron scattering measurements were performed on the WOMBAT high-intensity powder diffractometer at the OPAL reactor, Australian Centre for Neutron Scattering, ANSTO \cite{studer_wombat_2006}. A neutron wavelength of \SI{1.54}{\angstrom} was selected using a Ge(113) monochromator. The instrument is equipped with a large curved two-dimensional position-sensitive detector comprising eight multi-wire \ce{^3He} segments spanning approximately $120^\circ$ in $2\theta$, enabling efficient collection of weak diffuse signals. A radial collimator was employed to suppress scattering from the sample environment. Single crystals of \ce{MAPbBr3} and \ce{FAPbBr3} were mounted on an Eulerian cradle and aligned with the $(hk0)$ plane in the scattering geometry. Data were collected by rotating the sample about the vertical axis in $0.2^\circ$ steps with \SI{1}{\minute} exposures per frame, yielding a reciprocal space coverage of approximately 1.1 to \SI{7.5}{\per\angstrom} in $Q$. Temperature-dependent measurements were carried out using a closed-cycle cryostat at 100, 200, and \SI{300}{\kelvin} for \ce{MAPbBr3} and at 100 and \SI{200}{\kelvin} for \ce{FAPbBr3}. The raw data were processed using the LAMP software package to normalise, background-subtract, and reconstruct flat sections of reciprocal space.

\subsection*{Molecular Dynamics Simulations}
Machine-learning molecular dynamics simulations were used to support the assignment of diffuse-scattering features and to connect reciprocal-space signatures with real-space octahedral-tilt correlations. Different force fields were used depending on composition: a previously developed unified MACE force field for the MA/FA-containing iodide--bromide compositions, a standalone MACE force field for \ce{MAPbCl3}, an Allegro force field for \ce{CsPbBr3}, and a neuroevolution potential (NEP) force field for \ce{CsPbCl3}. The static and dynamic structure factors, $S(\mathbf{q})$ and $S(\mathbf{q},\omega)$, were computed from the resulting MD trajectories using the gpuscatter package \cite{gpuscatter}.

\subsubsection*{\ce{MAPbI3}, \ce{MAPbBr3}, \ce{FAPbI3}, and \ce{FAPbBr3}}
For the MA/FA-containing iodide--bromide compositions, including \ce{MAPbI3}, \ce{MAPbBr3}, \ce{FAPbI3}, and \ce{FAPbBr3}, we used the previously developed unified MACE force field for the \ce{(Cs/MA/FA)Pb(I/Br)3} chemical space \cite{liang_fapi_twin}. This model was trained on DFT-labelled structures spanning mixed A-site and mixed-halide compositions, with endpoint-family structures included for \ce{CsPbX3}, \ce{MAPbX3}, and \ce{FAPbX3} \cite{liang_perovskite_2025}. The final production model used a \SI{5}{\angstrom} radial cutoff, 64 feature channels, and message equivariance up to $L=0$. The final training set contained 6000 DFT-labelled structures and used a 90:10 train--test split.

Trajectories were propagated in LAMMPS through the MACE-MLIAP interface with cuEquivariance acceleration. Simulations used $18\times18\times18$ pseudocubic supercells, a \SI{1}{\femto\second} timestep, and Nosé--Hoover thermostat/barostat control with damping constants of \SI{100}{\femto\second} and \SI{200}{\femto\second}, respectively. For each composition and temperature, eight independent initial configurations were sampled. Constant-temperature simulations were equilibrated for \SI{200}{\pico\second} in the NpT ensemble, followed by \SI{100}{\pico\second} of NVT equilibration and a final \SI{200}{\pico\second} NVE structure-collection stage. Snapshots for real-space structural analysis were saved every \SI{1}{\pico\second}, while trajectories for calculating $S(\mathbf{q},\omega)$ were saved every \SI{50}{\femto\second}.

\subsubsection*{\ce{MAPbCl3}}
\ce{MAPbCl3} lies outside the iodide--bromide chemical space of the unified MACE model and was therefore described using a separate standalone MACE force field. This \ce{MAPbCl3} model was trained using the same MACE architecture and hyperparameters as the unified production model, including a \SI{5}{\angstrom} radial cutoff, 64 feature channels, and message equivariance up to $L=0$. The training dataset was generated from on-the-fly VASP molecular dynamics, followed by static DFT single-point recalculation of selected structures to obtain energies, forces, and stresses. The on-the-fly sampling protocol and DFT settings were kept consistent with those used for the unified MACE force field. Sampling was performed at 50, 125, 175, 250, and \SI{400}{\kelvin}. Because \ce{MAPbCl3} has a simpler structural and phase-transition sequence than the iodide--bromide chemical space, a smaller final dataset of 293 DFT-labelled frames was used.

Production MD simulations for \ce{MAPbCl3} followed the same protocol as for the unified MACE simulations, except that $16\times16\times16$ pseudocubic supercells were used.

To connect to the single crystal X-ray diffuse scattering measurements, the dynamical structure factor $S(\mathbf{q},E)$ was calculated from the MD trajectories using the pynamic structure factor (psf) package \cite{psf2023}. For each material and initial configuration, the \SI{0.5}{\nano\second} trajectories were divided into $10$ blocks of \SI{50}{\pico\second}. $S(\mathbf{q},E)$ was then averaged over these blocks and (in fractional reciprocal lattice units) over the two initial configurations. We extracted $S(\mathbf{q},E)$ for 0$\leq$ H,K,L $\leq$5 and expanded these values to the whole plane by mirroring in the coordinate axes. $q$-dependent atomic form factors, approximated as a sum of Gaussians, were used as described in the SI of \cite{weadock_nature_2023} and implemented in psf \cite{psf2023}. Based on these MD trajectories, real-space structural dynamics analysis was performed with the PDynA package, in order to extract octahedral tilt angles, tilt-correlation lengths, tilt-fluctuation amplitudes and real-space nanodomain morphology \cite{liang_structural_2023}.       

\subsubsection*{\ce{CsPbI3}}

For \ce{CsPbI3}, we used the publicly available MD trajectories of Baldwin et al.\ \cite{baldwin_dynamic_2024}. Briefly, these were generated with a machine-learned interatomic potential based on the Atomic Cluster Expansion (ACE), fitted by Bayesian linear regression and trained through the Hyper Active Learning (HAL) active-learning framework against DFT reference data computed with the FHI-aims code at the local density approximation level. Production simulations were run in the NpT ensemble with a \SI{4}{\femto\second} timestep on supercells of 69{,}120 atoms ($24\times24\times24$ pseudocubic unit cells), sampling configurations every \SI{200}{\femto\second} over a \SI{1}{\nano\second} window after \SI{1}{\nano\second} of equilibration. Full methodological details are given in Ref.~\cite{baldwin_dynamic_2024}.

\subsubsection*{ \ce{CsPbBr3}}

The MD simulations of \ce{CsPbBr3} used an Allegro \cite{allegro,allegro2} MLIP. DFT training data was gathered using the VASP on-the-fly structure selection process \cite{vasp_original1,vasp_original2}, where a Gaussian approximation potential (GAP)-style potential is fit on-the-fly. Training data are extracted based on a Bayesian error prediction \cite{vaspmlff_original}. These NpT runs used $2\times2\times2$ pseudo-cubic supercells at 500, 350, 240, 150 K. The r$^{2}$SCAN exchange-correlation functional \cite{r2scan_original1}, a plane wave basis set cut-off energy of \SI{500}{\eV}, and a $2\times2\times2$ $\Gamma$-centred \textit{k}-point grid was adopted. The energy threshold for electronic convergence was set to 10$^{-5}\,$eV, and a Gaussian smearing with a width of 50$\,$meV was applied for the smearing of electronic band occupancy. To avoid issues related to the incomplete basis set when large volume changes occur during the on-the-fly MD run, an additional single-point DFT calculation was performed on all structures, and these recalculated forces, energies and stresses made up the final training, validation and test sets. 

The model used a 7 \si{\angstrom} cutoff and the rest of the model parameters were consistent with our previous Allegro models for \ce{(MA/FA)PbBr3} \cite{dubajic_dynamic_2025}. The training and validation sets contained 2078 and 260 and were shuffled after each epoch. The model was trained with the Adam optimizer in pytorch \cite{pytorch_2019} for $\sim$2000 epochs, using a batch size of 5 and a learning rate of 0.001. A loss function with a 1:1:1 weighing of the Allegro per atom mean squared energy, force and stress terms, respectively, was used. 
The Allegro MD of \ce{CsPbBr3} simulations were performed LAMMPS \cite{LAMMPS}, using the pair\_allegro patch \cite{pair_allegro} using $20\times20\times20$ supercells of pseudo-cubic unit cells and a \SI{4}{\femto\second} timestep. The simulation cells were first equilibrated using fixed-shape NPT dynamics at \SI{400}{\kelvin} for \SI{1}{\nano\second}. From this equilibrated state we ran \SI{2}{\nano\second} of NVE dynamics, dumping atomic positions every \SI{100}{\femto\second}, from which $S(\mathbf{q},\omega)$ was extracted.

\subsubsection*{\ce{CsPbCl3}}

MD simulations of \ce{CsPbCl3} were performed with the GPUMD package \cite{gpumd} using a previously developed NEP \cite{Fransson2023}. The simulations were carried out for the cubic phase at \SI{300}{\kelvin} in the NVT ensemble with a \SI{1}{\femto\second} timestep, using lattice parameters obtained from separate NpT simulations. A $48\times48\times48$ supercell of the cubic primitive cell was used, corresponding to 552{,}960 atoms. The static structure factor, weighted by X-ray form factors, was computed from the MD trajectories using the dynasor package \cite{dynasor1,dynasor2}.

The static structure factor was also calculated within a harmonic approximation. Because the cubic phase exhibits imaginary phonon modes, an effective harmonic model was first constructed by fitting force and displacement data \cite{Hellman2013} using the hiphive package \cite{hiphive}. This effective harmonic model was then used to generate thermally displaced snapshots representative of the harmonic ensemble, from which the structure factor was computed using dynasor. These calculations used a $16\times16\times16$ supercell, corresponding to 20{,}480 atoms.

The simulated structure factors obtained from MD and from the effective harmonic model agree well, indicating that a large part of the diffuse scattering arises from normal phonons. The rods connecting the $R$ and $M$ points are also reproduced, demonstrating that these features originate from soft phonon modes that can be captured by an effective harmonic model. This is consistent with previous findings for the oxide perovskite \ce{BaZrO3} \cite{BZO}, where excess diffuse scattering intensity at the $R$ point was explained in terms of a soft phonon mode and effective harmonic models.

\subsection*{Differential scanning calorimetry}

Differential scanning calorimetry (DSC) measurements were carried out using a Netzsch DSC 204 F1 Phoenix instrument in an Al crucible under a \ce{N2} atmosphere. Experimental conditions, including temperature range and ramp rate, were adjusted to the sample-specific phase behaviour. Cooling to temperatures below $-40\,^{\circ}$C was performed using liquid nitrogen, whereas above $-40\,^{\circ}$C cold nitrogen gas was used; because the cooling mode changes at this temperature, the signal-to-noise ratio also changes across this range. For \ce{FAPbI3} powders, the filled DSC pan was first annealed above 150 $^{\circ}$C to convert any residual $\delta$-phase \ce{FAPbI3} to the $\alpha$-phase prior to measurement.

For \ce{MAPbI3} the same single crystal was used for all measurements. For the 3 K/min rate for \ce{MAPbI3} we measured from 353.15 K (80 $^{\circ}$C) to  123.15 K (-150 $^{\circ}$C). For the slower rates we had to adjust the temperature range. For 0.4 K/min we measured from 353.15 K (80 $^{\circ}$C) to 123.15 K (-150 $^{\circ}$C). For 0.1 K/min we measured from 338.15 K (65 $^{\circ}$C) to 313.15 K (40 $^{\circ}$C). For 0.01 K/min we measured from 333.15 K (60 $^{\circ}$C) to 323.15 K (50 $^{\circ}$C).

\subsection*{Hyperspectral photoluminescence microscopy}

Wide-field hyperspectral photoluminescence (PL) measurements were performed on a Photon etc.\ IMA system, using a ZEISS Plan-Neofluar objective (numerical aperture 0.75, $\times 63$ magnification) for all measurements. To limit degradation from oxygen and humidity, samples were stored in a nitrogen-filled glovebox until immediately before measurement. For the temperature-dependent measurements, samples were fixed with silver paste to the sample stage of an Oxford HiRes Microstat. No cryogen was used, so the stage was heated rather than cooled; the temperature and ramp rate were regulated by a proportional-integral-derivative (PID) controller through an in-house Python program, which allowed two distinct heating rates to be applied while maintaining accurate temperature control. A reference sample was used to calibrate the apparatus and to determine the post-processing parameters needed to correct image distortion introduced by the optics in the detection path. Because thermal expansion of the cryostat head shifts the focal plane, and to mitigate chromatic aberration, the sample $z$-position (focus) was adjusted automatically for each central wavelength and each temperature using a previously recorded calibration; this focus correction was applied concurrently with acquisition through an automatic $z$-stage, so that all measured intensities remain reliable.

A \SI{640}{\nano\meter} picosecond pulsed laser was used as the excitation source, operating at a repetition rate of \SI{1}{\kilo\hertz} and a fluence of \SI{0.34}{\micro\joule\per\square\centi\meter} with a top-hat profile ($150 \times 150~\mu\text{m}^2$). This low repetition rate was chosen so that no laser-induced increase or decrease in PL intensity was observed, a common indicator of laser-induced damage, and the PL signal remained stable even at the higher measurement temperatures. To ensure clean and consistent surface responses, single crystals were cleaved immediately before each measurement. PL was collected from a field of view of $85 \times 85~\mu\text{m}^2$ confirmed to be free of observable surface heterogeneities. Although hyperspectral microscopy provides a PL spectrum at every pixel, the response was integrated over the entire field of view to improve the signal-to-noise ratio, which is particularly important at higher temperatures where the PL is weaker. Integrating in this way ensures that the measured PL reflects the intrinsic properties of the material and minimises contributions from surface morphology and contamination; PL near rough surfaces or cracks can be substantially altered, and, unlike macroscopic PL collected from a larger area with less control over surface quality, this approach provides a more reliable characterisation of the intrinsic PL of single crystals.

The excitation laser was separated from the PL signal using a high-quality \SI{640}{\nano\meter} Semrock dichroic mirror. The emitted light was dispersed spectrally by a volume Bragg grating and detected by a PIMAX 4 emICCD camera ($1024 \times 1024$ pixels, \SI{13}{\micro\meter} pixel size).

\section*{Author Contributions}

M.D. conceived of the project and performed the data analysis and visualisation.
X.L., J.K. and E.F. constructed the machine learning force field and performed the molecular dynamics (MD) simulations under the supervision of A.W. and J.W.

M.D., X.L., J.K and E.F. performed the diffuse scattering simulations using the MD trajectories.  
P.H. performed differential scanning calorimetry measurements and contributed to the growth of \ce{FAPbI3} single crystals.
B.M.G. contributed to the growth of \ce{FAPbI3} single crystals.
M.D., M.v.Z., Y.L. and T.A.S. carried out single crystal X-ray diffraction measurements.
C.O. laser-cut crystals for single crystal X-ray diffraction.
Q.G. grew \ce{MAPbCl3} single crystals.
J.M. synthesised \ce{FAPbI3} powder. 
G.T.U. grew \ce{FAPbBr3} and triple-cation single crystals.
K.S.B. and I.G. grew \ce{CsPbBr3} and \ce{CsPbCl3} single crystals.
C.L. grew \ce{MAPbBr_{1.5}Cl_{1.5}} single crystals.
M.S. provided \ce{MAPbBr3} single crystals.
J.J., S.A., M.P.N., T.W., S.P.B., A.H.-B., M.K., H.J.S., A.W. and S.D.S. contributed to the interpretation of the results.
M.D. wrote the manuscript with the assistance of S.D.S.
S.D.S supervised the project. 
All authors contributed to reviewing and editing the manuscript.  

\section*{Conflicts of Interest}
S.D.S. is a cofounder of Swift Solar Inc and Clarity Sensors Limited.

\section*{Data Availability}
The data that support the findings of this study is available to download at the University of Cambridge Apollo Repository [DOI to be added at acceptance].

\section*{Code availability}
The Python and Matlab code that support the findings of this study are publicly available at GitHub [link to be added at acceptance]. The main software tools used are also openly available: rspace3d, for reciprocal-space reconstruction of single crystal X-ray diffraction data (\url{https://github.com/dubajicmilos/rspace3d}), and gpuscatter, for computing the static and dynamic structure factors $S(\mathbf{q})$ and $S(\mathbf{q},\omega)$ from molecular dynamics trajectories (\url{https://github.com/dubajicmilos/gpuscatter}), both developed in this work, together with the PDynA package used for real-space octahedral-tilt analysis (\url{https://github.com/WMD-group/PDynA}).

\section*{Acknowledgments}
\begin{sloppypar}
 
The authors would like to acknowledge the ANSTO beam time received on WOMBAT.
The authors thank the staff from the Mark Wainwright Analytical Centre at UNSW Sydney for the X-ray and DSC measurements.
Via our membership of the UK's HEC Materials Chemistry Consortium, which is funded by EPSRC (EP/X035859/1), this work also used the ARCHER2 UK National Supercomputing Service (http://www.archer2.ac.uk).
The training of the machine-learned force fields was enabled by the Berzelius resource provided by the Knut and Alice Wallenberg Foundation at the National Supercomputer Centre. We acknowledge the National Academic Infrastructure for Supercomputing in Sweden (NAISS) partially funded by the Swedish Research Council through grant agreement no. 2022-06725 for awarding this project access to the LUMI supercomputer, owned by the EuroHPC Joint Undertaking, hosted by CSC (Finland) and the LUMI consortium.
M.D. acknowledges UKRI guarantee funding for Marie Skłodowska-Curie Actions Postdoctoral Fellowships $2022$ \((\text{EP}/\text{Y}024648/1)\) and support by AINSE Limited through a PGRA award.
T.A.S. acknowledges funding from   EPSRC   Cambridge   NanoDTC, \((\text{EP}/S022953/1)\).
X.L. acknowledges support from the InnoCore program of hydro*studio (MSIT: 1.260005.01) at UNIST, dedicated to advanced postdoctoral training.
Y.L. acknowledges financial support from the Engineering and Physical Sciences Research Council (EPSRC, EP/V06164X/1).
J. K. acknowledges support from the Swedish Research Council (VR) program 2021-00486.
The authors thank the Leverhulme Trust  (\(\text{RPG-}2021\text{-}191)\) for funding.
We acknowledge DESY (Hamburg, Germany), a member of the Helmholtz Association HGF, for the provision of experimental facilities. Parts of this research were carried out at PETRA III, and we would like to thank Fernando Igoa and Katharina Köhler for assistance in using the P21.1 beamline. Some of the equipment we utilised on P21.1 was funded by the German Federal Ministry of Education and Research (BMBF) under grant 05K22RF1. M.P.N. recognizes support from the UNSW Scientia Program and an ARC DECRA Fellowship (Grant No. DE230100382). At Northwestern this study was supported by the Defense Threat Reduction Agency (DTRA) through the Interaction of Ionizing Radiation with Matter University Research Alliance (IIRM URA), under contract HDTRA1 20 2 0002 (K.S.B).

S.D.S. acknowledges the Royal Society and Tata Group (grant no. UF150033, URF/R/221026). We thank Diamond Light Source for access and support in the use of beamline I19-1 (proposal CY41632-1), I19-2 (proposal CY36628-1) and I15 (proposal CY38508-2). This work is supported by a European Research Council grant (VAPOURISE, 101169608). Views and opinions expressed are however those of the authors only and do not necessarily reflect those of the European Union or the European Research Council Executive Agency. Neither the European Union nor the granting authority can be held responsible for them. 

\end{sloppypar}

\newpage

\bibliography{references.bib,custom-refs.bib}

\end{document}